\documentclass{article}

\usepackage{arxiv}

\usepackage[utf8]{inputenc} 
\usepackage[T1]{fontenc}    
\usepackage{hyperref}       
\usepackage{url}            
\usepackage{booktabs}       
\usepackage{amsfonts}       
\usepackage{nicefrac}       
\usepackage{microtype}      
\usepackage{lipsum}
\usepackage{xcolor}
\usepackage{amsmath}
\usepackage{graphicx}
\usepackage{algorithm}
\usepackage{algorithmic}
\usepackage{mathtools}
\graphicspath{ {./images/} }

\DeclareMathOperator*{\argmin}{arg\,min}
\DeclareMathOperator{\sgn}{sgn}

\title{Enhancing Dynamic Mode Decomposition Workflow with In-Situ Visualization and Data Compression}

\author{
    Gabriel F. Barros \\
  Dept. of Civil Engineering\\
  COPPE/Federal University of Rio de Janeiro \\
  P.O. Box 68506, RJ 21945-970, Rio de Janeiro, Brazil \\
  \texttt{gabriel.barros@coc.ufrj.br} \\
     \And
  Malú Grave \\
  Dept. of Civil Engineering\\
  COPPE/Federal University of Rio de Janeiro \\
  P.O. Box 68506, RJ 21945-970, Rio de Janeiro, Brazil \\
  \texttt{malugrave@nacad.ufrj.br} \\
   \And
 José J. Camata \\
 Dept. of Computer Science \\
 Federal University of Juiz de Fora \\
 36036-330, Juiz de Fora, Brazil \\
 \texttt{camata@ice.ufjf.br}
    \And
 Alvaro L.G.A. Coutinho \\
  Dept. of Civil Engineering\\
  COPPE/Federal University of Rio de Janeiro \\
  P.O. Box 68506, RJ 21945-970, Rio de Janeiro, Brazil \\
  \texttt{alvaro@nacad.ufrj.br} \\
}

\begin{document}
\maketitle
\begin{abstract}

Modern computational science and engineering applications are being improved by the advances in scientific machine learning. Data-driven methods such as Dynamic Mode Decomposition (DMD) can extract coherent structures from spatio-temporal data generated from dynamical systems and infer different scenarios for said systems. The spatio-temporal data comes as snapshots containing spatial information for each time instant. In modern engineering applications, the generation of high-dimensional snapshots can be time and/or resource-demanding. In the present study, we consider two strategies for enhancing DMD workflow in large numerical simulations: (i) snapshots compression to relieve disk pressure; (ii) the use of in situ visualization images to reconstruct the dynamics (or part of) in runtime. We evaluate our approaches with two 3D fluid dynamics simulations and consider DMD to reconstruct the solutions. Results reveal that snapshot compression considerably reduces the required disk space. We have observed that lossy compression reduces storage by almost $50\%$ with low relative errors in the signal reconstructions and other quantities of interest. We also extend our analysis to data generated on-the-fly, using in-situ visualization tools to generate image files of our state vectors during runtime. On large simulations, the generation of snapshots may be slow enough to use batch algorithms for inference. Streaming DMD takes advantage of the incremental SVD algorithm and updates the modes with the arrival of each new snapshot. We use streaming DMD to reconstruct the dynamics from in-situ generated images. We show that this process is efficient, and the reconstructed dynamics are accurate.

\end{abstract}

\section{Introduction}

\par
The use of data-driven methods in engineering and scientific applications has been growing considerably in the past decades with the advent of modern techniques, dedicated hardware, and data availability. Extracting information from data to improve decision-making in many different fields is one of the main pillars of the modern scientific machine learning framework. The current state-of-the-art of engineering applications - which comprises the use of expensive simulations originating from numerical methods to approximate complex and (usually) nonlinear transient partial differential equations (PDEs) - can also benefit by learning from data \cite{Brunton2019}. 
Subjects such as dynamical systems model discovery \cite{Brunton2016, Kaiser2018}, model order reduction \cite{Benner2021, Peherstorfer2018, Carlberg2011, Hesthaven2015} and control \cite{Proctor2016, Kutz2016} have been widely explored in recent years with data-driven approaches and applied to a diverse range of applications such as neuroscience, turbulence, epidemiology and finance \cite{Brunton2019}.
\par 
However, since data fuel these approaches, one must be cautious regarding I/O operations. Nowadays, the disparity between computing and I/O speed \cite{foster2021online} makes it very difficult for large simulations to output to disk a large amount of data to be analyzed later. This is particularly true when seeking accurate solutions for the numerical discretization of PDEs, where the state vectors obtained from the nonlinear systems are likely high-dimensional (i.e., the use of fine meshes on finite element simulations) and require substantial storage space when saved into disk. Consequently, data should be analyzed online (while the simulation is running), or data reduction computations should be performed online to reduce the amount of data saved to disk. One possible way to mitigate this issue is to consider lossy data compression \cite{Sayood2012} on the resulting simulation data. This approach can dramatically improve storage space by reducing the state vectors floating point precision. The use of data compression is widely considered to improve computational performance and model accuracy in computational science and engineering \cite{Diffenderfer2019, ainsworth2018multilevel, ainsworth2019multilevel, ainsworth2019multilevel2, ainsworth2020multilevel, foster2021online, jin2022accelerating}, and in machine learning \cite{Bratko2006, Littlestone1986, Sculley2006, Ghahramani2015} applications. Another issue that arises from time-demanding simulations or time-critical problems is that the complete data is available only after the end of the simulations. The use of on-the-fly algorithms for data-driven methods able to update the model with the addition of new snapshots has been explored in the literature (see \cite{Phalippou2020, matsumoto2017onthefly, Peherstorfer2015}). In this case, with real-time snapshots generation, one can analyze and steer numerical parameters during runtime \cite{Silva2020}, perform real-time diagnostics on time-variant systems\cite{Zhang2019} and real-time predictions \cite{Alfatlawi2020}. Recently, in \cite{newberry2021software} it was proposed an immersive simulation software framework that describes the relationship between a PDE  solver and simulation software components that enable in-situ visualization and data analysis, computational steering, and modification of physical, geometric, and solver parameters. All this has tremendous implications for parallel applications. The result is a new parallel program structure, known as \emph{Online Data Analysis and Reduction} (ODAR), that has implications for the co-design of extreme-scale computers \cite{foster2021online}. This new motif enables more efficient design space exploration, reducing the gap between simulations, data analysis, and insight extraction. 

\par 
In the present study, we present enhancements to the Dynamic Mode Decomposition (DMD) workflow, a data-driven method based on snapshots. The snapshots are outputs from parallel simulations, the \texttt{Simulate} phase in the ODAR terminology \cite{foster2021online}. DMD stacks snapshots into a matrix and maps the nonlinear dynamics using a linear operator, that is, the \texttt{Analyze} ODAR phase. In this paper, we will introduce the \texttt{Reduce} phase (data compression) in the DMD workflow and also discuss how in-situ visualization strategies can be embedded in the workflow. We consider different data compression strategies applied to snapshots and evaluate the accuracy of the results. We assess the storage efficiency of lossless and lossy compression methods with different accuracy thresholds. Then, we evaluate the use of DMD on the resulting data to assess how the compression affects the reconstruction/prediction accuracy. Two 3D fluid dynamics problems with multiple degrees of freedom per node are considered to illustrate the efficiency of data compression. The first case simulates a lock-exchange simulation of a particle-laden flow, where fluids and sediments interact and generate the forces driving the fluid motion. The following problem is a bubble rising example, where the difference in fluid densities is responsible for the dynamics. Both simulations are parallel and usually present considerable computational costs due to moving fronts, fine meshes, and high complexity due to the number of variables and their interactions, making them ideal for data compression approaches. Data is consequently decompressed, pre-processed, and the respective dynamic system approximation is computed using standard DMD \cite{Kutz2016, Schmid2010}. Results are evaluated in terms of relative error, and other relevant physical properties are verified for all tested cases. Since both simulations demand a significant amount of time to complete, we also apply streaming DMD \cite{Zhang2019, Alfatlawi2020} on the data generated using \textit{in situ} visualization tools (more precisely, Paraview Catalyst \cite{ayachit2015paraview, ayachit2021catalyst}) at runtime. In this case, images generated by Paraview Catalyst scripts are input for streaming DMD, where each snapshot is assessed and used to update the modes if it contains relevant information. The relative errors in the results, that is, the error between approximated pixel values and original pixel values from the images, are also compared using different metrics.

The remainder of this paper is structured as follows: Section \ref{sec:DMD} briefly describes DMD and its variants used in this work. In Section \ref{sec:data}, we discuss the main data compression and \textit{in situ} data visualization techniques. We also describe our approach to reducing storage requirements on complex simulations and extracting snapshots using data visualization software at runtime. In Section \ref{sec:examples} we evaluate our strategies on the aforementioned examples. In Section \ref{sec:conclusions} we present our conclusions and future works.

\section{Dynamic Mode Decomposition}
\label{sec:DMD}
Dynamic Mode Decomposition (DMD) is a data-driven method that aims to extract coherent structures from spatio-temporal data without prior information about the system. DMD can be used on experimental and numerical data and capture nonlinear phenomena using a linear approach. DMD was proposed for the fluid dynamics context  \cite{Rowley2009,Schmid2010}, being extended to a wide range of applications such as  climate \cite{Kutz2016}, epidemiology \cite{Proctor2015, Barros2021, Viguerie2022compart}, tumor growth \cite{Viguerie2022tumor}, biomechanics \cite{Calmet2020}, additive manufacturing \cite{Viguerie2022compart}, especially in structure extraction from data and control-oriented methods. 

\subsection{DMD Structure}
\par 
DMD acts as follows: given a set of vectors containing spatially distributed measurements in time, the first step is to build a snapshots matrix $\mathbf{Y}$ by stacking the vectors horizontally, such that 
\begin{equation}
                \mathbf{Y} = \left[
                \begin{array}{cccc}
                \vrule & \vrule &        & \vrule \\
                \mathbf{y}^h_0  & \mathbf{y}^h_1  & \ldots & \mathbf{y}^h_m    \\
                \vrule & \vrule &        & \vrule
                \end{array}
                \right]
\end{equation}
where $\mathbf{y}^h_k$ are the observations made for time instants $k = 0, 1, \dotsc, m$, where $m+1$ is the total number of observations. The observations here are the nodal values obtained from parallel finite element simulations, expressed by the superscript $^h$. The snapshots matrix can be decomposed into $\mathbf{Y}_1$ and $\mathbf{Y}_2$ such that $\mathbf{Y}_1 = [\mathbf{y}^h_0 \dotsc \mathbf{y}^h_{m-1}]  \in \mathbb{R}^{n \times m}$ and $\mathbf{Y}_2 = [\mathbf{y}^h_1 \dotsc \mathbf{y}^h_m]  \in \mathbb{R}^{n \times m}$. 
That said, a linear mapping can be made between both matrices, that is, 
\begin{equation}
    \mathbf{Y}_2 = \mathbf{A}\mathbf{Y}_1
\end{equation} 
where $\mathbf{A}$ contains the dynamical properties that transform dataset $\mathbf{Y}_1$ into dataset $\mathbf{Y}_2$. In the standard procedure (often called as \textit{exact DMD} \cite{bopdmd}), the linear mapping $\mathbf{A}$ is obtained by solving the optimization problem

\begin{equation}
    \mathbf{A} = \argmin_{\mathbf{A}} || \mathbf{Y}_2 - \mathbf{A}\mathbf{Y}_1||_F = \mathbf{Y}_2\mathbf{Y}_1^\dagger
\end{equation}
where $||\cdot||_F$ denotes the Frobenius norm and the superscript $\dagger$ represents the Moore-Penrose pseudoinverse. This approach, however, is not computationally advisable since explicitly computing the Moore-Penrose pseudoinverse may bring excessive computational cost to the algorithm due to the curse of dimensionality \cite{bopdmd}. The pseudoinverse may also be ill-conditioned, resulting in large errors. A more efficient approach is to compute the Singular Value Decomposition (SVD) of $\mathbf{Y}_1$, truncate into the first $r$ singular vectors that hold the most part of the variance of $\mathbf{Y}_1$ and then compute the matrix $\mathbf{\tilde{A}}$, a $r\times r$ projection of $\mathbf{A}$. This is obtained by  
\begin{equation}
    \mathbf{\tilde{A}} = \mathbf{U}^T_r\mathbf{A}\mathbf{U}_r = \mathbf{U}^T_r\mathbf{Y}_2\mathbf{V}_r\mathbf{\Sigma}_r^{-1}
\end{equation}
where $\mathbf{Y}_1 = \mathbf{U}\mathbf{\Sigma}\mathbf{V}^T$ and the subscript $r$ indicates the truncation of the first $r$ singular values/vectors. We highlight that $\mathbf{\tilde{A}}$ is unitarily similar to $\mathbf{A}$. Then, the eigendecomposition of $\mathbf{\tilde{A}}$ is essential for computing the DMD modes $\mathbf{\Psi}$, that is
\begin{center}
	\begin{equation}
	\mathbf{\Psi} = \mathbf{Y}_2\mathbf{V}_r\mathbf{\Sigma}^{-1}_r\mathbf{W},
	\end{equation}
\end{center}
where the matrix $\mathbf{W}$ contains the eigenvectors $\mathbf{\phi}_j$ of $\mathbf{\tilde{A}}$. Finally, the signal approximation can be written as:
\begin{equation}\label{DMDExpansion}
    \mathbf{y}^h(t)  \approx  \tilde{\mathbf{y}}^h(t) =  \mathbf{\Psi}\exp(\mathbf{\Omega}_{eig}t) \mathbf{b},
\end{equation}
where $\mathbf{b}$ is a vector containing the initial conditions projected onto the DMD modes such that $\mathbf{b} = \mathbf{\Psi^{\dagger}}\mathbf{y}^h_0$. The matrix $\mathbf{\Omega}_{eig}$ is diagonal and its entries are the continuous eigenvalues $\omega_i = \ln(\lambda_i)/\Delta t_o$, 
where $\Delta t_o$ is the time step size between the snapshots and $\lambda_i$ is the \textit{i}th eigenvalue of $\mathbf{\tilde{A}}$. 
\par 
Finally, the choice of $r$ in the DMD context is subjective \cite{Taira2017} and usually done by trial and error. Although some techniques are employed in specific contexts such as noisy or experimental data \cite{Kutz2016, Gavish2014, Donoho1995}, a good starting point is to evaluate the retained variance in data by considering a hard threshold technique \cite{Kutz2016} such that,
\begin{equation}
	\label{eq:relative_energy}
	\kappa = 1 - \dfrac{\sum_{i=1}^{r}\sigma_{i}^{2}}{\sum_{i=1}^{m}\sigma_{i}^{2}} \leq \tau,
\end{equation}    
where $\tau$ is a defined tolerance threshold (usually $10^{-6}$). This method indicates that more than $100(1-\tau)\%$ of the variance in the data is preserved by the approximation.
\par 
Also, it is important to mention that since the whole dataset is considered on the construction of the basis, we call this algorithm standard DMD\cite{Kutz2016}. In this case, data is already available and preprocessed for computation. A different approach, also discussed in section \ref{sec:svd} of this study, is the streaming DMD\cite{Zhang2019, Alfatlawi2020}, where data is available one snapshot at a time and the DMD basis is initialized and updated with each new snapshot. Standard and streaming DMD are named according to the classification used in \cite{Zhang2019}.

\subsection{Singular Value Decomposition} 
\label{sec:svd}
The core algebraic operation of standard DMD is the Singular Value Decomposition (SVD). The SVD factorization applies to any complex $n \times m$ matrix. By definition, the squared singular values of a given matrix $\mathbf{M}$ are the eigenvalues of $\mathbf{M}\mathbf{M}^T$ or $\mathbf{M}^T\mathbf{M}$, where the eigenvectors of the first product lead to the left singular vectors $\mathbf{U}$ and the eigenvectors of the latter product are the right singular vectors $\mathbf{V}$. However, more efficient algorithms were developed to compute the SVD factorization\cite{Dongarra2018}. The seminal work of Sirovich \cite{Sirovich1987} introduced the Method of Snapshots, a  milestone in efficiently computing SVD-based reduced-order models. Various algorithms for SVD were then developed, such as randomized factorizations \cite{Halko2011, Erichson2019} and real-time decompositions \cite{Phalippou2020, matsumoto2017onthefly, Brand2002}. In this study, we consider two different SVD decomposition algorithms for different situations: the randomized SVD (rSVD) \cite{Erichson2019} and the incremental SVD (iSVD) algorithms \cite{matsumoto2017onthefly}. 

\par 
The rSVD is a non-deterministic algorithm that computes the near-optimal low-rank approximation of a given large dataset with good efficiency. The snapshots matrix is post-multiplied by a random test matrix of dimensions $\mathbf{R} \in \mathbb{R}^{m \times (r + p)}$ where $p$ is the number of columns chosen for oversampling, a technique that ensures that the resulting matrix $\mathbf{Z}$ spans the column space of the snapshots matrix, that is
\begin{equation}
    \mathbf{Z} = \mathbf{Y} \mathbf{R}.
\end{equation}

After computing $\mathbf{Z}$, an optional, although advisable, step is to submit the matrix to power iterations, that is, repeatedly updating $\mathbf{Z} = \mathbf{Y} (\mathbf{Y}^T\mathbf{Z})$ for $q$ times. This step, which can dramatically increase computational cost depending on the value of $q$, improves the quality of the basis by ensuring a more rapid decay of the singular values. The resulting matrix $\mathbf{Z}$ is then submitted to QR factorization to generate a low-dimensional orthonormal basis $\mathbf{Q}$ for the snapshots matrix. The snapshots matrix is projected onto the lower-dimensional basis, resulting in matrix $\mathbf{B} = \mathbf{Q}^{T}\mathbf{Y}$. The standard SVD routine is invoked for $\mathbf{B}$, returning $\mathbf{\hat{U}}$, $\mathbf{\hat{\Sigma}} = \mathbf{\Sigma}$ and $\mathbf{\hat{V}} = \mathbf{V}$. Aside from $\mathbf{\hat{U}}$, the matrices obtained from the SVD decomposition on the lower dimension are equivalent to the matrices obtained on the high dimensional snapshots matrices. The left singular vectors for the snapshots matrix are obtained by re-projecting the smaller dimensions singular vectors $\mathbf{U} = \mathbf{Q}\mathbf{\hat{U}}$. Reducing the dimensionality of the problem before extracting the singular values can improve the efficiency of the overall SVD routine in orders of magnitude \cite{Barros2020cilamce}. 

\par 
Apart from the rSVD, we considered the iSVD for streaming data. The iSVD is especially useful in real-time applications where DMD is applied for control or when simulations are excessively time-demanding, making data scarcely available in the simulation's early stages and progressively available. This data flow can enable insights from different stages of the dynamics instead of the usual standard DMD. In this study, we have implemented a slightly different version \cite{Oxberry2017} of the usual iSVD algorithm, initially proposed for Proper Orthogonal Decomposition (POD) applications. In this algorithm, for a given collection of snapshots $\mathbf{Y}$, a new snapshot $\mathbf{y}$ updates the singular values and vectors such that
\begin{equation}
    \left[\begin{array}{cc}
        \mathbf{Y} &  \mathbf{y}
    \end{array} \right] = 
    \left[\begin{array}{cc}
        \mathbf{U}_Y &  \mathbf{h}
    \end{array} \right] 
    \left[\begin{array}{cc}
        \mathbf{\Sigma}_Y &  \mathbf{l} \\
        \mathbf{0} &  g
    \end{array} \right] 
    \left[\begin{array}{cc}
        \mathbf{V}_Y &  \mathbf{0} \\
        \mathbf{0} &  1
    \end{array} \right]^T 
\end{equation}
where $\mathbf{U}_Y$, $\mathbf{\Sigma}_Y$, $\mathbf{V}_Y$ are the singular values and vectors of $\mathbf{Y}$, being $\mathbf{Y}$ the matrix containing the already existent snapshots during the increment of the new snapshot $\mathbf{y}$. If no snapshots are available, the matrices can be initialized as $\mathbf{\Sigma}_Y = [||\mathbf{y}||]$, $\mathbf{U}_Y = [\mathbf{y}/||\mathbf{y}||] $, $\mathbf{V}_Y = [1]$, being $\mathbf{y}$ the first snapshot available and $||\cdot||$ the $\mathcal{L}^2$ norm. Vectors $\mathbf{l} = \mathbf{U}_Y^T\mathbf{y}$ and $\mathbf{h} = (\mathbf{y} - \mathbf{U}\mathbf{l})/g$ and scalar $g = \sqrt{\mathbf{y}^T\mathbf{y} - \mathbf{l}^T\mathbf{l}}$ are responsible for updating the snapshots basis. It is important to mention that not all snapshots are needed to improve the basis. If $g < \epsilon_{SVD}$, where $\epsilon_{SVD}$ is a predefined threshold, the snapshot does not influence the new basis by any means. In this study, we consider $\epsilon_{SVD} = 10^{-15}$. It is also important to truncate $\mathbf{U}_Y$, $\mathbf{\Sigma}_Y$, $\mathbf{V}_Y$ to preserve the rank $r$. Also, an orthogonalization check is advisable on $\mathbf{U}_Y$ at each increment, invoking a QR decomposition on $\mathbf{U}_Y$ in case of loss of orthogonality of the basis. At the end of the acquisition of the snapshots, the generated SVD matrices will preserve the first $k$ more relevant vectors of the included snapshots on the system. The standard DMD approach is employed for the examples where all data is available at the time of the computation of the DMD modes. In this case, the rSVD algorithm is invoked. However, when considering \textit{in situ} visualization for real-time data generation, the streaming DMD is preferred, and the iSVD is considered. Both standard and streaming DMD algorithms are presented respectively in Algorithms \ref{algdmd} and \ref{algstream}.


\begin{algorithm}
	\caption{Reconstruction of dynamical systems using the standard DMD method}
	\begin{algorithmic}
		\STATE \textbf{INPUT:} Snapshots matrix $\mathbf{Y} = \{\mathbf{y}_0, \hdots , \mathbf{y}_m\}$
		\STATE \textbf{OUTPUT:} Signal reconstruction $\mathbf{\tilde{y}}(t) \approx \mathbf{y}(t)$
		\STATE 1: Set $\mathbf{Y}_1 = \{\mathbf{y}_0, \hdots , \mathbf{y}_{m-1}\}$ and $\mathbf{Y}_2 = \{\mathbf{y}_1, \hdots , \mathbf{y}_m\}$.
		\STATE 2: Compute the SVD of  $\mathbf{Y}_1$,  $\mathbf{Y}_1 = \mathbf{U\Sigma V^T}$.
		\STATE 3: Define the truncation rank $r$ 
		\STATE 4: Compute $\mathbf{\tilde{A}} \coloneqq \mathbf{U}_r^T\mathbf{Y}_2\mathbf{V}_r\mathbf{\Sigma}_r^{-1}$.
		\STATE 5: Compute eigenvalues and eigenvectors of $\mathbf{\tilde{A}W = W\Lambda}$.
		\STATE 6: Compute continuous eigenvalues $\Omega$ from discrete eigenvalues $\Lambda$, $\omega_i = \ln(\lambda_i) / \Delta t_i$, being $\Delta t_i = t_{i+1} - t_i$.
		\STATE 7: Set $\mathbf{\Psi}^{DMD} = \mathbf{Y}_2 \mathbf{V}_r \mathbf{\Sigma}^{-1}_r \mathbf{W}$.
		\STATE 8: Compute vector $\mathbf{b}$, $\mathbf{b} = (\mathbf{\Psi}^{DMD})^\dagger \mathbf{y_0}$.
		\STATE 9: Reconstruct approximation $\mathbf{\tilde{y}}(t)$, $\mathbf{\tilde{y}}(t) = \sum_{i = 1}^{k}b_i \psi_i\exp(\omega_i t)$
	\end{algorithmic}
	\label{algdmd}
\end{algorithm}

\begin{algorithm}
	\caption{Reconstruction of dynamical systems using the Streaming DMD method}
	\begin{algorithmic}
		\STATE \textbf{INPUT:} Independent snapshots $\{\mathbf{y}_i\}$, number of DMD modes $r$, $\epsilon_{SVD}$, SVD matrices $\mathbf{U}, \mathbf{V}, \mathbf{\Sigma}$ if already initialized.
		\STATE \textbf{OUTPUT:} Signal reconstruction $\mathbf{\tilde{y}}(t) \approx \mathbf{y}(t)$
		\IF{$i = 0$}
		    \STATE 1: $\{\mathbf{y}_i\} = \{\mathbf{y}_0\}$. Matrices $\mathbf{U}$, $\mathbf{V}$ and $\mathbf{\Sigma}$ must be initialized.
		    \STATE 2: Initialize $\mathbf{U} = \mathbf{U}_Y \gets [\mathbf{y}_0/||\mathbf{y}_0||]$
		    \STATE 3: Initialize $\mathbf{\Sigma} = \mathbf{\Sigma}_Y \gets [||\mathbf{y}_0||]$
		    \STATE 4: Initialize $\mathbf{V} = \mathbf{V}_Y \gets [1]$
		    \STATE 5: Initialize $\mathbf{Y} = [\mathbf{y_i}]$
		    \STATE 6: Initialize rank counter $r_{count}$
		\ELSE 
		    \STATE 7: Compute vector $\mathbf{l} = \mathbf{U}^T_Y \mathbf{y}_i$
		    \STATE 8: Compute scalar $g = \sqrt{\mathbf{y}_i^T \mathbf{y}_i - \mathbf{l}^T\mathbf{l}}$
		    \STATE 9: Compute vector $\mathbf{h} = (\mathbf{y}_i - \mathbf{Ul})/g$
		    \IF{$g > \epsilon_{SVD}$}
		        \STATE 10: Update $\mathbf{Y} \gets [\mathbf{Y} \hspace{0.15cm} \mathbf{y}_i]$
		        \STATE 11: Apply SVD on $\mathbf{Q} = \left[\begin{array}{cc}
                                            \mathbf{\Sigma}_Y &  \mathbf{l} \\
                                            \mathbf{0} &  g
                                        \end{array} \right] =  \mathbf{U'\Sigma' V'^T}$.
                \IF {$r_{count} \leq r$}
                    \STATE 12: Update $\mathbf{U} \gets \left[\begin{array}{cc}
                                                \mathbf{U}_Y &  \mathbf{h}
                                            \end{array} \right] \mathbf{U'}$
                    \STATE 13: Update $\mathbf{V} \gets \left[\begin{array}{cc}
                                                \mathbf{V}_Y &  \mathbf{0} \\
                                                \mathbf{0} &  1
                                            \end{array} \right] \mathbf{V}'^T$
                    \STATE 14: Update $\mathbf{\Sigma} \gets \mathbf{\Sigma'}$
                \ELSE 
                    \STATE 15: Update $\mathbf{U} \gets \mathbf{U}_Y \mathbf{U'}[:,:r]$
                    \STATE 16: Update $\mathbf{V} \gets \mathbf{V}_Y  \mathbf{V}'^T[:, :r]$
                    \STATE 17: Update $\mathbf{\Sigma} \gets \mathbf{\Sigma'}[:,:r]$
                \ENDIF
		    \ENDIF 
		\ENDIF
		\STATE 18: Returns $\mathbf{U}$, $\mathbf{V}$, $\mathbf{\Sigma}$ and $\mathbf{Y}$. 
		Inference proceeds identically as seen in lines 4-9 of Algorithm \ref{algdmd}.   
	\end{algorithmic}
	\label{algstream}
\end{algorithm}

\section{Data Reduction and streaming techniques}
\label{sec:data}

The traditional large-scale scientific simulation workflow consists of defining the physical problem of interest, its domain, and its appropriate boundary and initial conditions as an initial phase. Later, the numerical solution is obtained through an appropriate PDE solver. In this same phase, raw data is written to disk. Finally, we perform data visualization and analysis to determine the usefulness of the solution's data. With the advent of exascale computing, large-scale scientific applications are likely to consume and produce massive amounts of data. It has been shown that this post hoc analysis paradigm is no longer scalable \cite{foster2021online}. The reason for this lack of scalability is primarily due to slow disk I/O speed compared to the rate at which data is produced, high storage requirements, and prohibitive data transfer costs. As a result, only a subset of time steps of the simulation can typically be stored on disk for future analysis.
Thus, the traditional scientific workflow must be rethought to circumvent several issues related to managing large data sets generated by that applications. Two proposals emerge to mitigate this bottleneck in this context: in-situ data visualization and data reduction.

\begin{figure}[ht!]
	\centering
	\includegraphics[scale=0.75]{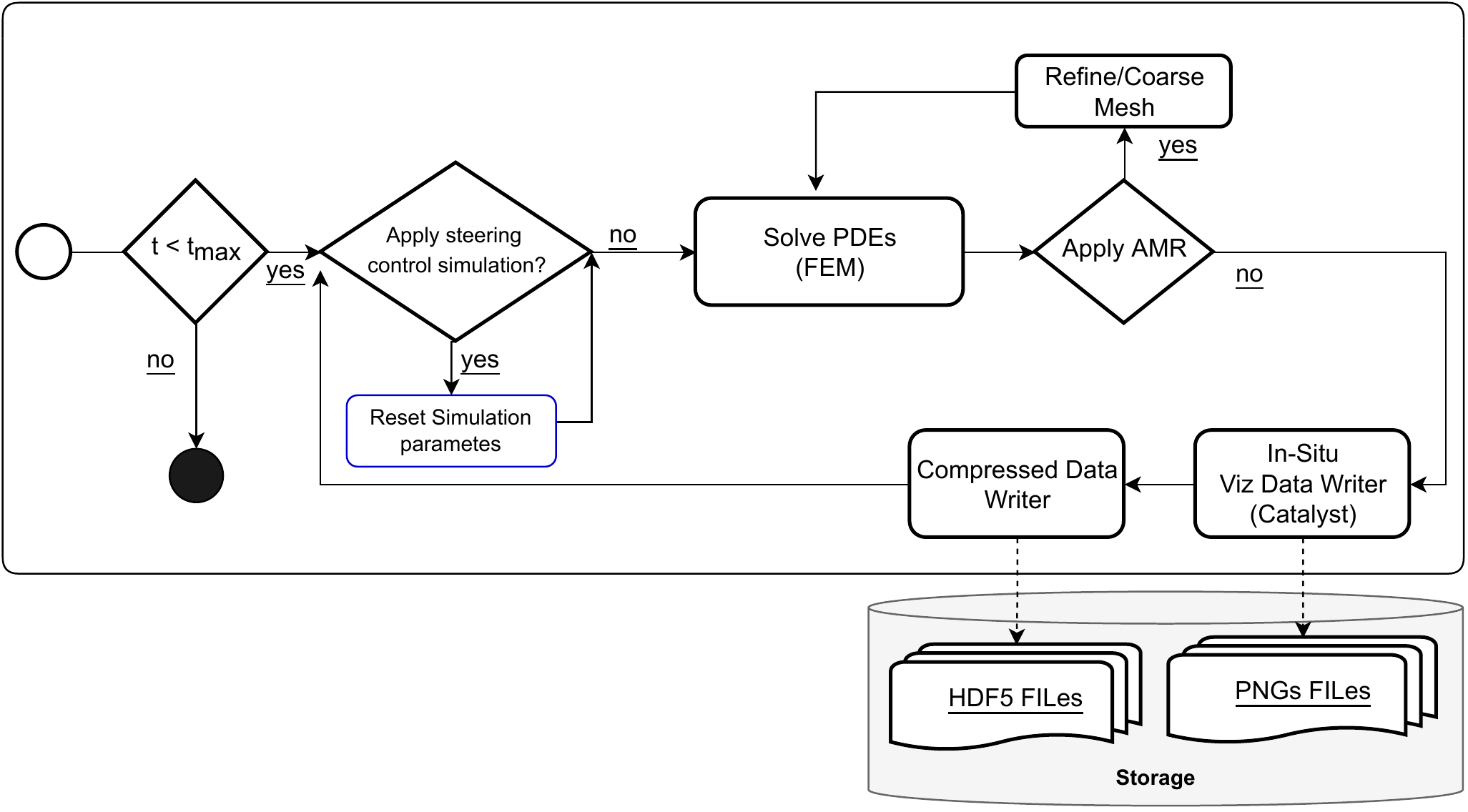}
	\caption{Multiphysics simulation workflow. The simulation time interval is $[0, t_{max}]$.}
	\label{fig:flowchart}
\end{figure}

Figure \ref{fig:flowchart} illustrates our proposed multi-physics numerical simulation workflow which is augmented by: \textit{(i)} an in-situ visualization and data analysis component and \textit{(ii)} a compressed Data Writer component. Moreover, two other components complete the multi-physics workflow: user-steering control and adaptive mesh refinement and coarsening (AMR/C).  

Our multiphysics simulation workflow is built on top of \texttt{libMesh}, a C++ FEM open-source software library for parallel adaptive finite element applications \cite{libMeshPaper}. This library provides native support for AMR/C on general unstructured meshes. In addition, libMesh uses high-quality third-part software such as PETSc and Trilinos for the solution of linear and non-linear systems on both serial and parallel platforms. Although libMesh provides many tools for mesh-based solvers, applications built upon the library need to implement some finite element kernels according to the physics of the problem of interest and the numerical formulation adopted. The user-steering control component uses DfAnalyzer, which is a solution based on provenance data to extract and relate strategic simulation data in transit from multiple data for online queries \cite{Silva2020}. As a result, our multiphysics simulator has been used to solve several real problems such as dense granular flows \cite{Gesenhues_2019}, turbidity currents \cite{Camata2018}, and bubble dynamics \cite{Grave2020}. Particularly, in Camata et al. \cite{Camata2018} the user-steering component is used to monitor sediments appearance from turbidity currents simulations. The tools enable monitoring of the sediment appearance at runtime and steer the simulation based on the solver convergence and visual information on the sediment deposits, enhancing the analytical power of such simulations.

\subsection{In-situ visualization and data analysis}
In recent years, in-situ visualization and data analysis simulation workflows have gained considerable attention from the Computational Science and Engineering, and High-Performance Computing communities. Fundamentally, these workflows include a co-processing stage in the simulation by integrating pre- and post-processing steps. One of the emerging technologies is the ability to perform runtime in-situ visualization and data analysis, reducing the amount of data written to disk and mitigating bottlenecks in data transfers. Several frameworks have been developed to provide in-situ visualization capabilities, such as, ParaView Catalyst \cite{ayachit2015paraview, ayachit2021catalyst} and Visit-libsim \cite{HPV:VisIt}. Other initiatives emerged to integrate different in-situ visualization and data analysis infrastructure over the same interface like Sensei \cite{sensei2016} and Alpine/ZFP  \cite{alpine}.
 
For the in-situ image data writer component, we integrate our simulation workflow with ParaView Catalyst \cite{ayachit2015paraview, ayachit2021catalyst}.  
ParaView Catalyst is an in-situ data processing visualization developed upon ParaView \cite{ParaView}. To enable Catalyst on a parallel simulation code, we have implemented a Catalyst adaptor with \texttt{Initialize}, \texttt{CoProcess}, and \texttt{Finalize} methods that invoke pipelines defined in Python scripts to extract simulation data and/or to generate visualizations. \texttt{Initialize} method starts Catalyst in a proper state in a simulation run. This method is invoked once per simulation run and before \texttt{CoProcess} method invocations. It reads the analyses pipelines defined in Python scripts generated by the ParaView UI application. The \texttt{CoProcess} method is called in time steps relevant for data analysis and visualization. PNG image files can be generated according to pipelines defined in the initialization process in these steps. Then, the \texttt{Finalize} method is responsible for releasing memory allocated during the simulation run and cleaning up the Catalyst state. 


\subsection{Data Compression}

The introduction of compression into a scientific workflow can benefit storage limitations, I/O bottlenecks, and bandwidth. In general, the compressors can be split into two categories, lossless and lossy data compressors. Generic lossless compression algorithms such as Gzip and  BZIP have shown less effective for float-point data sets generated by scientific applications \cite{lindstrom_fast_2006}. Although there are lossless compressors designed for numerical data \cite{lindstrom_fast_2006,burtscher_fpc_2009, claggett_spdp_2018}, they have limited compression rates due to high precision representation. In this way, the least significant bits of this data have random characteristics, disfavoring the prediction of recurring values. However, leading bits tend to favor the development of a new class of error-bounded lossy compression algorithms \cite{lindstrom2014,di_fast_2016}. 

One of these emerging methods is the ZFP library for compressed floating-point values \cite{zfp2014}. It is recognized as one of the most effective high-speed lossy ﬂoating-point data compressors. ZFP operates on $d$-dimensional arrays ($d \in [1,4]$) by partitioning them into blocks of $4^d$ values. Each block is compressed/decompressed entirely independently from all other blocks. The scheme uses orthogonal block transform and embedded coding ideas to achieve a high compression ratio. Therefore,  ZFP users could control a block's quality and compressed size by modifying several library parameters. There are five predefined compression modes: expert, fixed-rate, fixed-precision, fixed-accuracy, and reversible. The first four modes allow a higher compression ratio with controllable loss of accuracy, while the last one performs lossless compression. Details about these compression modes can be found in  \cite{zfp2019}.


All data generated by the multiphysics kernel during the time loop presented in Figure \ref{fig:flowchart} are written in XDMF/HDF5 format \footnote{\url{https://xdmf.org/index.php/Main_Page}}. The eXtensible Data Model and Format (XDMF) was created to standardize the scientific data between High-Performance Computing (HPC) codes. XDMF uses XML  to store heavy raw data and metadata that describes the Data Model. Heavy data can be stored in either Hierarchical Data Format (HDF5) \cite{hdf5} or binary files. In this work, all raw data are written in HDF5 files. HDF5 is a high-performance data model software library widely used by the scientific community. It was designed for flexible and efficient I/O, high volume, and complex heterogeneous data.

The benefits of data compression can be obtained in HDF5 through user-provided dynamically loaded filters. HDF5 has the concept of a filter pipeline, which is just a series of operations performed on data. The library provides flexibility to use various data compression filters on individual HDF5 datasets. Compressed data is stored in chunks and automatically decompressed by the library and filter plugin when a chunk is accessed. The primary compression filters included in the library are GZIP and SZIP. 
However, HDF5 also allows third-party filters such as LZO, BZIP2, LZF, LZ4, FPZIP. All filters mentioned before apply a lossless data compression. Some of the filters available for lossy compression are SZ, FPZIP, and ZFP. The complete list of all filters can be found at the HDF Group website \footnote{\url{https://portal.hdfgroup.org/display/support/Registered+Filter+Plugins}}.

\subsection{DMD Workflow}
\begin{figure}[ht!]
	\centering
	\includegraphics[scale=0.90]{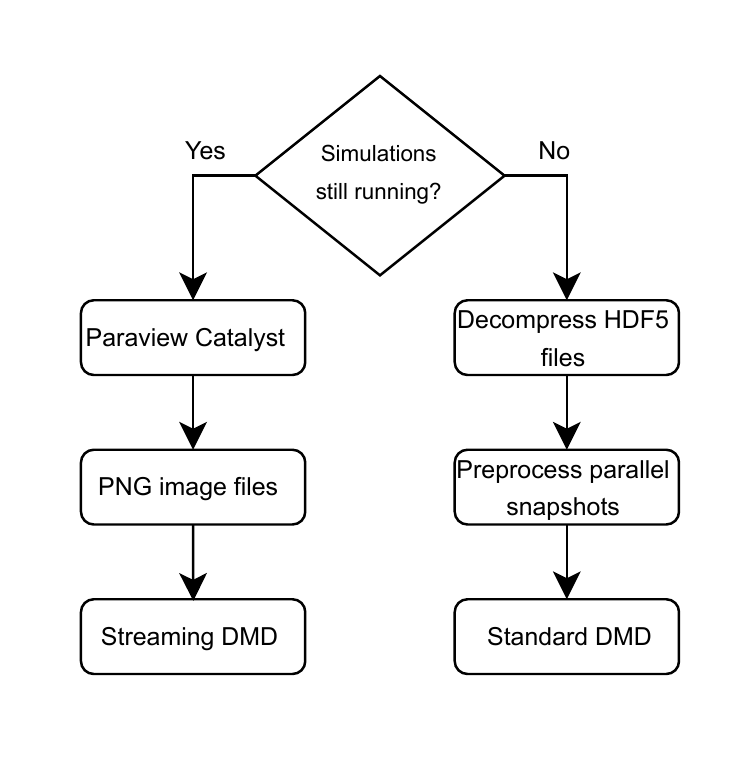}
	\caption{DMD Workflow: Left, Streaming DMD; Right Standard DMD.}
	\label{fig:online_phase}
\end{figure}

Data is generated and exported to disk by the numerical simulation code as shown in Figure \ref{fig:flowchart}. In this section, we explore the DMD workflow, the step where DMD uses the generated data as input for processing. Although standard DMD is the most common approach, for large simulations, given that the available data may be scarce, on-the-fly algorithms can be employed for control or short temporal predictions \cite{matsumoto2017onthefly}. For simulations that may take days or weeks to complete, one can use the generated data as input for reconstructing the dynamics, and, as new data is provided, the computed basis may be updated with new information. Instead of transferring large output files containing all the information related to the mesh and vector/scalar fields, in this study, we take advantage of Paraview Catalyst to create PNG image files of a view of the current simulation state to be input in runtime into the streaming DMD algorithm \cite{Zhang2019, Alfatlawi2020}, integrating the DMD workflow to the simulation workflow. One positive aspect of PNG files is that they are considerably smaller than the complete simulation data of a given time step, making data transfer easier and simpler. In this sense, instead of using nodal values for DMD processing, the pixels from PNGs can be treated as spatio-temporal data and used as input for DMD. Figure \ref{fig:online_phase} illustrates the DMD workflow. In this study, we consider two different input files: compressed HDF5 files and PNG files. The compressed HDF5 files are considered available as a batch (containing all the generated data) after the simulation is ended. The PNG files are available during the simulation, being generated on-the-fly via Paraview Catalyst. The workflow starts with the arrival of new data for both cases. The workflow left branch illustrates the DMD processing for streaming data. In this case, the simulations are still running, and data is being generated on-the-fly. The in-situ visualization tools are invoked to generate a PNG image file for the state vector on that given instant. If the simulation involves a 3D domain, care should be taken to generate the images. Usually, it is advisable to use a properly placed 2D view. The iSVD algorithm is responsible for updating the DMD basis for each PNG image file generated at runtime. Streaming DMD is then used to describe the system's dynamics or infer future states.

The right workflow branch in Figure \ref{fig:online_phase} describes the standard DMD procedure using compressed data. In this case, the simulation is completed, and all the output files are created using compressed HDF5 files and are ready to be collected and preprocessed. Data is read from the output files, each instant at a time. Snapshots at this point require preprocessing partitioned data available on the HDF5 files, given that the finite element simulations are running in parallel. This preprocessing is only applied to the HDF5 files, and processing PNG image files do not require this step. The idea is to remove communication nodes and to number the nodes in a way that all the snapshots have the correct spatio-temporal information for each spatial coordinate on the mesh without duplicates. 


Given that the data is correctly assembled on the snapshots matrix, the next step is to apply DMD to the data. On the compressed HDF5 files, standard DMD using the rSVD routine is invoked. Streaming DMD using the iSVD algorithm is preferred for PNG files, and the basis gets updated as new PNG files arrive. Then, each DMD model generates results for different purposes, explained in detail in section \ref{sec:examples}. Given the matrix dimensions for the present study, both algorithms can be evaluated in serial mode, given the fast processing power of the SVD routines and data dimensionality. In some cases, parallel SVD algorithms can be used to improve efficiency \cite{Dongarra2018, Maulik2021, Berry2005}. If the parallel SVD is considered, it is essential to mention that there is no correlation between the simulation and the SVD parallel data partitioning, meaning that the vector concatenation described should be done nevertheless.



\section{Numerical examples}
\label{sec:examples}

We now apply the presented strategies to two three-dimensional numerical examples. First, we test the results on a particle-driven gravity flow in a lock-exchange configuration \cite{Camata2018}. The simulation consists of filling a tank with both phases separated by a lock. The simulation starts when the lock between the phases - clear water and water with sediments - is released, and the difference between densities induces the fluid motion. The simulation is modeled by coupling the incompressible Navier-Stokes equation with the transport equation, representing the sediment flow. This complex multiphysics simulation contains multiple fields (i.e., velocities, pressure, sediments, deposition), that is, several degrees of freedom per node. This simulation is particularly suitable to test our data compression approach since the finite element solutions for each physics are expressed as floating-point arrays of dimension $n$, meaning that data compression should yield good performance. Also, since this is a multiphysics and highly nonlinear problem, it is expected that the simulation demands a large simulation time, being also interesting for the in-situ visualization snapshot generation approach. 

We also evaluate the results on a bubble rising simulation \cite{Grave2020}. The model involves coupling the incompressible Navier-Stokes equation with the convected level-set method\cite{Coupez2007, Ville2011, Grave2020}, an interface tracking method responsible for mapping each phase of the problem. The bubble rising problem is also a multiphysics and highly nonlinear problem. The simulation consists of modeling a gas bubble immersed in a liquid domain. The difference between the two densities drives the external force responsible for generating the fluid motion. Since the gas bubble is slightly less dense than the liquid, the bubble rises. 

Navier-Stokes and transport equations are approximated in this work by the residual-based variational multiscale finite element formulation \cite{Hughes2004, Rasthofer2017, Ahmed2017, Codina2018, Bazilevs2013,Guerra2013}. A staggered scheme advances in time the Navier-Stokes and transport equations. Non-linearities are handled by the Inexact-Newton method for both systems. We adopt the backward Euler scheme for the time integration of the Navier-Stokes equation and the second-order backward-difference (BDF2) method for the transport equation. Errors estimators are employed to control the adaptive mesh refinement and coarsening. We modify the standard workflow for both simulations, introducing the in-situ visualization and data compression tools explained in Section \ref{sec:data}. Thus, we can evaluate in this Section the effects of data compression on DMD online phase workflow and the reconstruction of the dynamics from PNG images with streaming DMD. Therefore, in the present paper, we run the simulations in parallel, generating compressed snapshots and PNG files at user-specified time intervals. The data reduction uses the ZFP algorithm with fixed-precision compression mode.


\subsection{Particle-driven gravity flow}
\par 
In this section, we apply DMD to reconstruct the solution of a simulation of a particle-driven gravity flow in a lock-exchange configuration. We consider two fluids with different densities due to the concentration of sediments. The difference between their densities is such that the Boussinesq hypothesis is valid. Moreover, particles in the heavy fluid have negligible inertia and are much smaller than the smallest length scales of the buoyancy-induced fluid motion. Thus, the dimensionless governing equations are, 

\begin{align}
		\nabla\cdot \mathbf{u} = 0,  \label{eq:mass_pdgf}\\
		\dfrac{\partial \mathbf{u}}{\partial t} + \mathbf{u}\cdot \nabla\mathbf{u} + \nabla p - \dfrac{1}{\sqrt{Gr}}\Delta \mathbf{u} =  c\mathbf{e}^g, \label{eq:momentum_pdgf}\\
		\dfrac{\partial c}{\partial t} + (\mathbf{u} + \mathbf{e}^gu_s)\cdot \nabla c - \dfrac{1}{Sc\sqrt{Gr}}\Delta c = 0.  \label{eq:transport_pdgf}
\end{align}
where $\mathbf{u}$ is the fluid velocity, $p$ is the pressure, $c$ the sediment concentration and $u_s$ is particle settling velocity  in the gravity direction $\mathbf{e}^g$. Equations \ref{eq:mass_pdgf}, \ref{eq:momentum_pdgf} and \ref{eq:transport_pdgf} are, respectively, the conservation of mass, momentum and sediment transport. 
Moreover, $Gr$ is the Grashof number that expresses the ratio between buoyancy and viscous effects, and $Sc$ is the Schmidt number that expresses the ratio between diffusion and viscous effects. Here we take $Gr=10^6$ and $Sc=1$. 

Essential and natural boundary conditions for Equations \ref{eq:mass_pdgf}-\ref{eq:momentum_pdgf} are $\mathbf{u} = \mathbf{g}$ on $\Gamma_g$ and
$(-p\mathbf{I} + \frac{1}{\sqrt{Gr}} \nabla \mathbf{u}) \cdot \mathbf{n} =\mathbf{h} \textrm{ on } \Gamma_h$ in which $\mathbf{g}$ is a prescribed velocity value on $\Gamma_g$ and $\mathbf{h}$ is the interaction traction force on $\Gamma_h$. Besides, a divergence-free initial condition for the velocity field must be specified. For Equation \ref{eq:transport_pdgf}, boundary conditions modeling the transport of particles are the Dirichlet boundary condition  $c = c_{in}$ on  $\Gamma_c$ which describes the quantity of sediment entering in the flow domain, no-flux boundary condition $(u_s\mathbf{g}c - \frac{1}{Sc\sqrt{Gr}} \nabla{c}) \cdot \mathbf{n} =\mathbf{0} \textrm{ on } \Gamma_h$ and convective flux condition $\frac{\partial c}{\partial t} - u_s \nabla c \cdot  \mathbf{n} = 0$ on ${\Gamma_b}$ with $\Gamma = \Gamma_{in} \cup \Gamma_h \cup \Gamma_b$ and  $\Gamma_{in} \cap \Gamma_h \cap \Gamma_b = 0$.

\begin{figure}[ht!]
    \centering
    \includegraphics[width=0.90\linewidth]{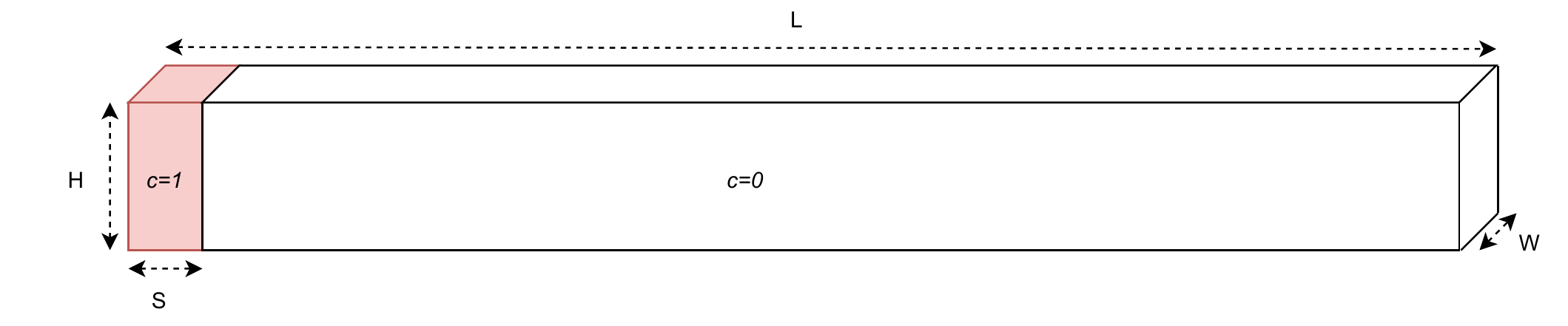}
    \caption{Geometry and initial conditions for the lock-exchange simulation.}
    \label{fig:ic_necker}
\end{figure}

\begin{figure}[ht!]
    \centering
    \includegraphics[width=0.95\linewidth]{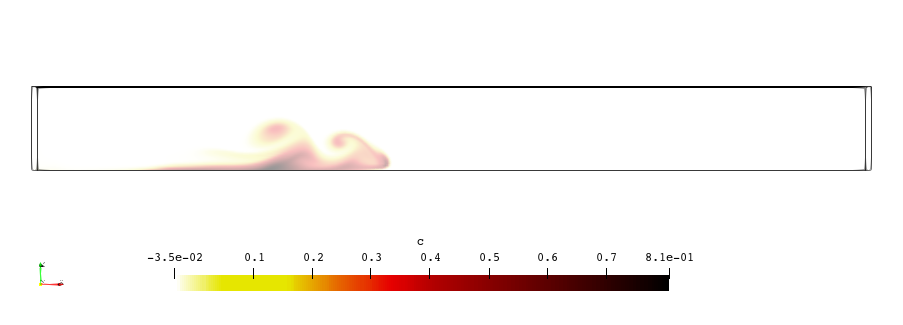}
    \caption{Sediment concentration snapshot at $t=10$ time units 
    -- the figure shows the sediment concentration 3D volume rendering.
    }
    \label{fig:lock_exchange_fom}
\end{figure}

Figure \ref{fig:ic_necker} illustrates the initial configuration for the simulation, which also depicts the dimensions of the domain $L=20$, $H=2$ and $W=0.1$. 
The simulation follows the lock-exchange configuration, where no injection of fluid or sediments is considered. Two fluids fill the domain, one with sediments and another with pure fluid, separated by a lock. The simulation starts with removing the lock ($S=0.75$), and the gradient of the concentration of sediments is responsible for generating the fluid motion. We use a hexahedral structured mesh with a $0.025$ grid spacing. The Navier-Stokes and sediment concentration equations linear systems of equations are solved in 16 cores using Block-Jacobi + GMRES(35) with local  ILU(0) preconditioning. GMRES tolerance is  $10^{-6}$. For the nonlinear solver, the tolerance is $10^{-3}$. The time step size is $0.001$, and the simulation runs until reaching the maximum time of 20 time units. XDMF/HDF5 raw data files are written every $10$ time steps, producing 2,000 snapshots. Catalyst generates PNG files for sediment concentration every $5$ time steps, resulting in 4,000 images. Figure \ref{fig:lock_exchange_fom} shows the concentration field at $t = 10$ time units.  The present results are in good agreement with earlier results \cite{Camata2018}.

\par 

The use of fine meshes leads to a large number of equations and, consequently, a significant increase in disk pressure to hold the snapshots. In this situation, data compression can decrease disk pressure and enable proper storage for the resulting simulation data. We assess the use of data compression on the simulation output data considering three different data compression strategies: lossless compression and lossy compression with accuracy thresholds of $10^{-6}$ and $10^{-3}$. 
Table \ref{tab:data_comp} shows for all test cases the resulting output size of HDF5 files for all snapshots.

\begin{table}[!htp]\centering
\caption{Storage results for the tested cases and time share spent for functions}
\begin{tabular}{l|cc|cccc}\toprule
                 & \multicolumn{2}{c}{\textbf{Storage Required}} & \multicolumn{4}{c}{\textbf{Share of time spent on}} 
                \\ & \textbf{XDMF/HDF5 (MB)} &  \textbf{Reduction (\%)}  &  \textbf{Solver} & \textbf{Write} & \textbf{Catalyst} & \textbf{Others}   \\
                \midrule
No compression & 35597                 &   --     &81.31\% &0.15\% &0.54\% &18.00\%                 \\
ZFP lossless & 29823                   & 16.15\%  &81.30\% &0.12\% &0.54\% &18.04\%                 \\
ZFP accuracy $10^{-6}$& 19100          & 46.15\%  &81.32\% &0.12\% &0.53\% &18.02\%                \\
ZFP accuracy $10^{-3}$ & 17841         & 49.67\%  &81.28\% &0.12\% &0.54\% &18.07\%                 \\
\bottomrule
\end{tabular}
\label{tab:data_comp}

\end{table}
\par 
Table \ref{tab:data_comp} shows that the lossless compression yields a $16\%$ reduction on the storage required for the generated data. Lossless compression indicates that the compressed data can be reverted to the original data without any information degradation. For lossy compression, we can define the desired accuracy threshold for floating point numbers, and, for larger thresholds, more reduction is observed. We notice that for a threshold of $10^{-6}$, the storage required for the HDF5 files is $46\%$ of the original data size, while for a threshold of $10^{-3}$, the compressed output files require almost half the storage needed for the original data. We also observe in Table \ref{tab:data_comp} that the times for the write operations and for generating the in-situ visualization images do not hamper the overall code performance. Moreover, data compression decreases the time spent writing data to persistent storage.

\par
\begin{figure}[ht!]
    \centering
    \includegraphics[width=0.49\linewidth]{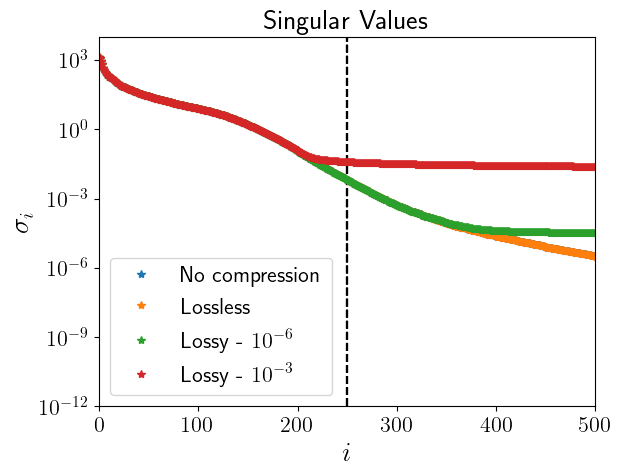}
    \caption{Singular values for the tested cases. The dashed line represents the number of $r = 250$ singular values and vectors selected for the DMD basis.}
    \label{fig:singval_necker}
\end{figure}

 
We now apply DMD on the sediment concentration data generated using all options in Table \ref{tab:data_comp}, although all fields were subjected to data compression. The \texttt{h5py} and \texttt{hdf5plugin} Python packages are responsible for adequately compressing and decompressing the data and reading the information written on the HDF5 files. The snapshots matrix is constructed, and the SVD decomposition of $\mathbf{Y_1}$ is computed. The DMD algorithm takes tens of seconds to run compared with the hours spent in the numerical simulation. Initially, we observe the singular values obtained from the SVD decomposition of $\mathbf{Y_1}$ on Figure \ref{fig:singval_necker}, where the curves show the singular values for the original data, data with lossless compression, and lossy compression with two different accuracies. We choose the $r = 250$ most relevant dynamic modes of the snapshots to reconstruct the solution. We note that the snapshot matrices generated from the simulations with no compression and lossless compression are identical, meaning that no information is lost during the compression process. The curves for data generated with lossy compression reveal that the decay for the singular values reaches a plateau at a given point. For accuracy equal to $10^{-6}$, most of the singular values are preserved compared to the lossless and no compression data. Given that the dashed line represents the truncation of the number of singular values for the DMD basis construction, the information lost would not affect the construction of the basis since the difference in the singular values lie ahead of the truncation threshold. For accuracy of $10^{-3}$, the truncation is more aggressive and affects the singular values before the basis truncation, changing the singular values preserved for the DMD basis.    

\par
\begin{figure}[ht!]
    \centering
    \includegraphics[width=0.95\linewidth]{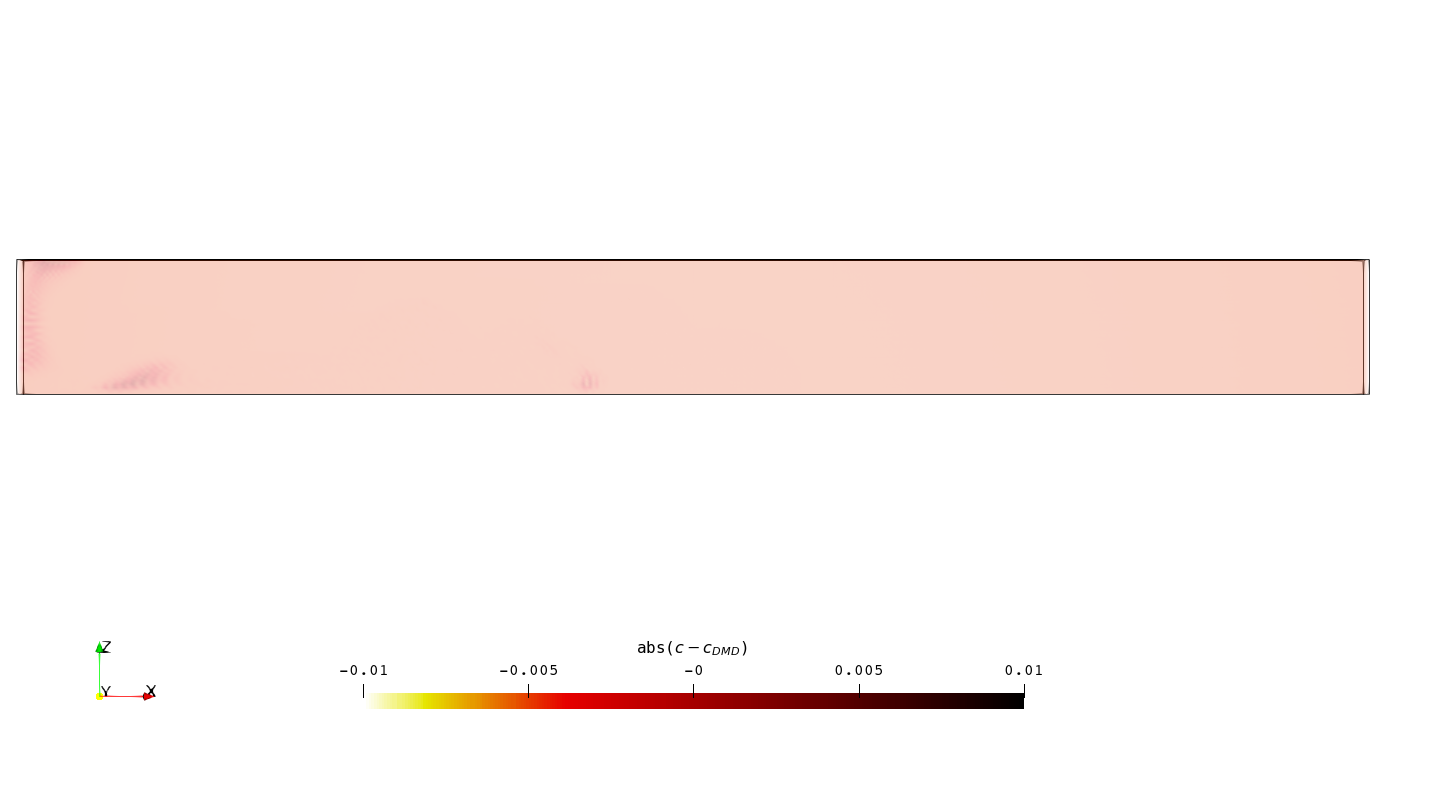}
    \caption{Absolute error between DMD reconstruction from lossy compressed data (threshold $10^{-6}$) and original snapshot for the sediment concentration at $t=10$ time units. }
    \label{fig:lock_exchange_abs_error}
\end{figure}

\begin{figure}[ht!]
    \centering
    \includegraphics[width=0.49\linewidth]{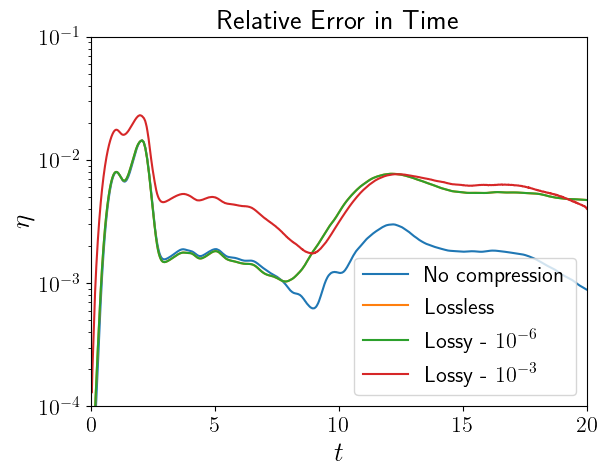}
    \caption{Relative error in time for the solution reconstruction.}
    \label{fig:rel_error_necker}
\end{figure}

 
After analyzing the singular values, we proceed to reconstruct the solution using standard DMD to approximate the results. Figure \ref{fig:lock_exchange_abs_error} shows the absolute error at $t = 10$ time units between the DMD reconstruction using lossy compression (threshold $10^{-6}$) and finite element solution. We can see that the DMD reconstruction at that time instant is of excellent quality. Figure \ref{fig:rel_error_necker} shows the relative error time history for DMD reconstructions and finite element solution, while Table \ref{tab:lockexchange_frobenius} shows the Frobenius relative error between the snapshots matrix and the matrix containing all the vectors approximated by DMD. The relative error is computed as $\eta = \dfrac{||\mathbf{y}^h_k - {{\mathbf{y}_k^h}_{DMD}||}}{{||{\mathbf{y}_k^h}||}}$ for $k = 0, 1, \dots, m$ snapshots and the Frobenius relative error is $\eta_F = \dfrac{||\mathbf{Y} - {{\mathbf{Y}}_{DMD}||_F}}{{||{\mathbf{Y}}||_F}}$, where $||\cdot||_F$ is the Frobenius norm. 

Observing Figure \ref{fig:rel_error_necker}, we note that the error magnitudes are of the same order in all cases. Again, the results for lossless compression and no compression are identical. For the lossy compression cases, we observe that the relative errors are slightly higher for the latter stages of the simulation. For $t < 8$s, we notice that the results for the lossy compression considering a threshold of $10^{-6}$ are very similar to those without lossy compression. It is also essential to notice that the relative error in time tends to fit the dynamics, in the sense that the instant where more significant errors are observed (i.e., $t = 2$s), the dynamics observed are more rapid, leading to more significant errors in comparison with the remainder of the simulation. We observe that the curves present similar behavior for all four cases, meaning there is no significant loss of physical accuracy when using data compression techniques. Further, Table \ref{tab:lockexchange_frobenius} shows the relative error in the Frobenius norm ($\eta_F$) for the original and approximated matrices. The relative Frobenius norm for all cases is of the same order of magnitude, but for the more aggressive data compression case, we note that $\eta_F$ is twice as larger. When running DMD, no significant additional computation time is observed for either compressing output files during simulation runtime or decompressing snapshots for DMD computation.

\begin{table}
    \centering
    \caption{Frobenius relative error between approximated and original snapshots for the particle-driven gravity flow considering $r=250$}
    \begin{tabular}{cc}\toprule 
        Test case & $\eta_F$ \\  
        \midrule
        No compression &  $5.28 \times 10^{-3}$ \\
        ZFP Lossless       &  $5.09 \times 10^{-3}$ \\
        ZFP accuracy ($10^{-6}$) & $5.31 \times  10^{-3}$ \\
        ZFP accuracy ($10^{-3}$) & $9.95 \times 10^{-3}$ \\
        \bottomrule
    \end{tabular}
    \label{tab:lockexchange_frobenius}
\end{table}

\begin{figure}[ht!]
    \centering
    \includegraphics[width=0.6\linewidth]{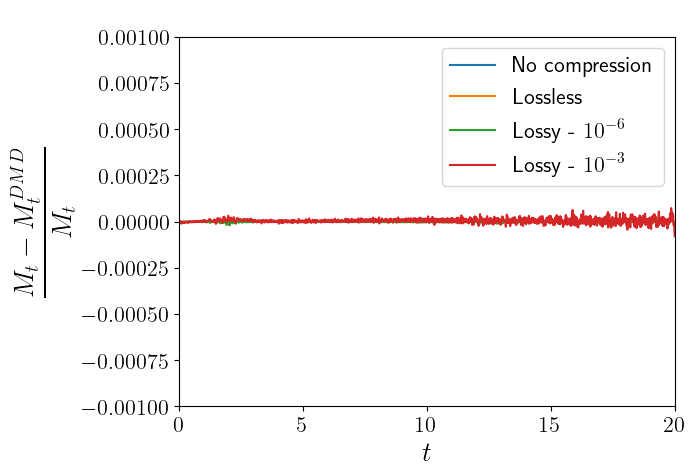}
    \caption{Time history of the relative sediment mass error for all cases. For this problem, $M_t = \int_\Omega c d\Omega$ and the superscript $^{DMD}$ refers to the same expression but taking the DMD reconstruction of $c$.}
    \label{fig:mass_necker}
\end{figure}

Another important quantity of interest to be assessed is sediment mass. That is, we need to verify if data compression influences quantities computed with the results of DMD reconstructions. We plot in Figure \ref{fig:mass_necker} the time history of the relative error between the sediment mass computed by the finite element simulation and the same quantity computed using the sediment concentration reconstructed by DMD ($r=250$) with uncompressed and compressed snapshots. The results show that DMD accurately reconstructs the sediment mass for the four cases. However, note that the lossy compression curve with an accuracy of $10^{-3}$ reveals the largest difference. 


\subsection{Bubble rising problem}
\begin{figure}[ht!]
\begin{center}
\includegraphics[width=0.3\textwidth]{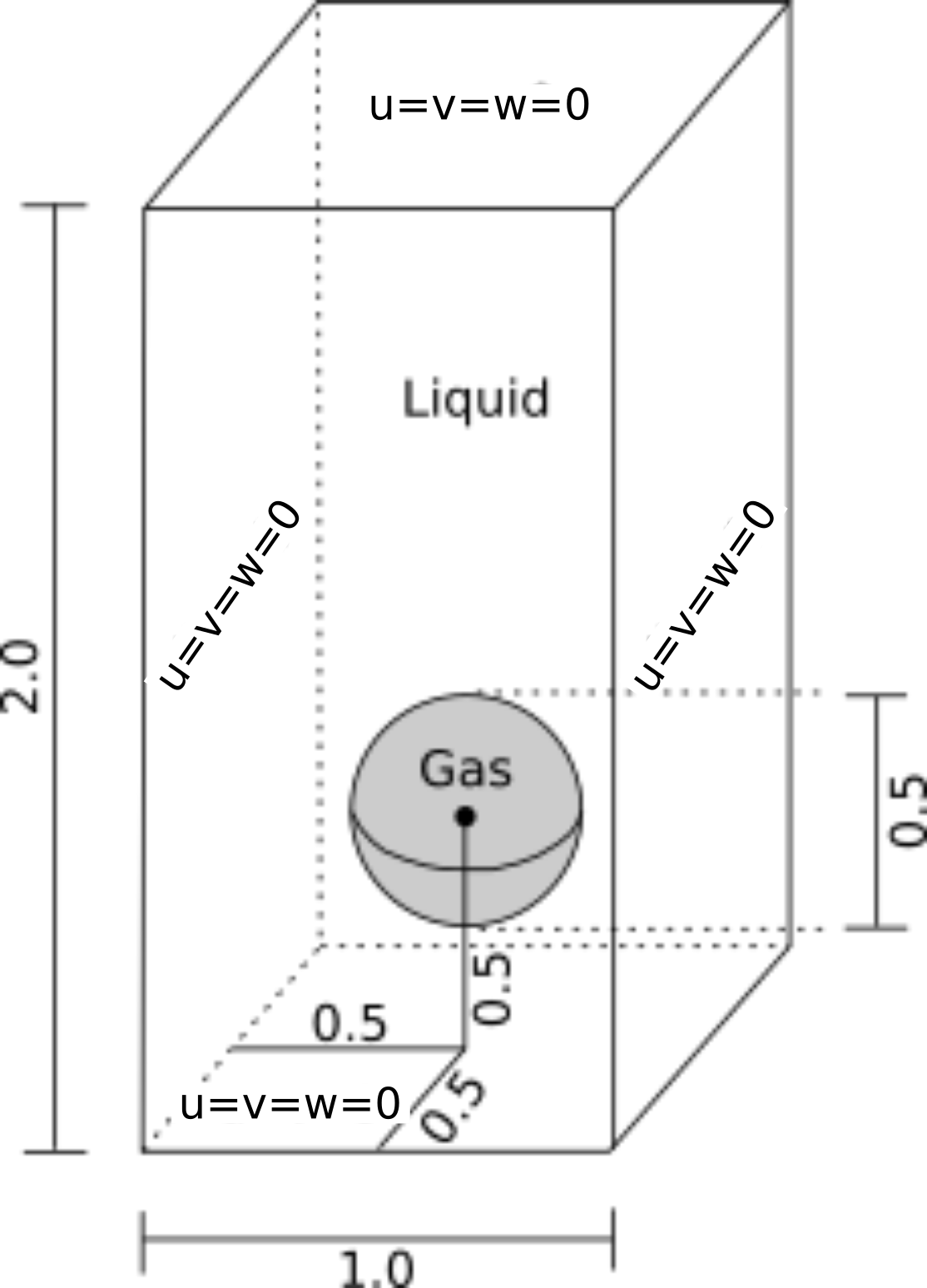}
\end{center}
\caption{Initial configuration and boundary conditions for the bubble rising problem.}
\label{bubble_bench3d}
\end{figure}

The next numerical test is a bubble rising 3D benchmark. In this simulation, we track the motion of a three-dimensional spherical gas bubble rising inside a liquid column. Figure \ref{bubble_bench3d} shows the initial configuration of the system, where the spherical bubble of radius $R = 0.25$ m is centered at $[0.5, 0.5,0.5]$ m in a $[1 \times 1 \times 2]$ m domain. All boundaries are prescribed with no-slip boundary conditions. Table \ref{tab:bubble} lists the parameters used for this simulation. The surface tension $\mathbf{F_{st}}$ plays a major role in bubble problems, and it is modeled using the Continuum Surface Model (CSF) \cite{brackbill1992continuum}.

\begin{table}
  \begin{center}
  \caption{Rising bubble data.}
   \label{tab:bubble}
    \begin{tabular}{c c c} 
      \hline
Computational domain & $1\times 1 \times 2$ & (m)
\\
Grid sizes & $0.100$ to $0.025$ & (m)
\\
Number of time steps & $240$ & (-)
\\
Time step & $0.0125$ & s
\\
Bubble radius & $0.25$ & m
\\
Initial bubble position & $(x,y) = (0.5, 0.5, 0.5)$ & m
\\
Liquid density & $1000$ & kg/$\text{m}^3$
\\
Liquid viscosity & $10$  & kg/(ms)
\\
Gas density &  $100$ & kg/$\text{m}^3$
\\
Gas viscosity & $1$ & kg/(ms)
\\
Surface tension &  $24.5$  & N/m
\\
Gravity &  $0.98$ &  m/$\text{s}^2$
\\
     \hline
    \end{tabular}
  \end{center}
  \end{table}
  


We write the governing equations in their dimensional form as 


\vspace{6pt}
\begin{center}
	\begin{equation}
	\begin{aligned}
	\nabla\cdot \mathbf{u} = 0,\\
	\rho\frac{\partial \mathbf{u}}{\partial t} +\rho\mathbf{u} \cdotp \nabla \mathbf{u} + \nabla p - \mu \nabla^2\mathbf{u} - \rho\mathbf{g} - \mathbf{F_{st}} = \mathbf{0},\\
	\frac{\partial \alpha}{\partial t} + (\mathbf{u} + \lambda \mathbf{U})\cdot \nabla \alpha - \lambda \sgn(\alpha) S = 0,
	\end{aligned}
	\label{level-set}
	\end{equation}
\end{center}
\vspace{6pt}

\noindent  where $\rho$ is the density, $\mu$ is the dynamic viscosity, $\mathbf{g}$ is the acceleration of gravity vector, $\alpha$ is the level-set function, $\lambda$ is a penalty constant, $\mathbf{U} = \sgn(\alpha) \dfrac{\nabla \alpha}{||\nabla \alpha||}$, and $S$ a function related to the level-set signed distance function.

For this simulation, we use an adaptive mesh. The AMR/C procedure uses a local error estimator to drive the refinement and coarsening procedure, considering the error of a finite element relative to its neighbor elements in the mesh. Depending on the calculated error, the original finite element is subdivided into smaller finite elements until a maximum level of refinement ($h_{max})$. Thus, we use a mesh, initially with $10 \times 10 \times 20$ cells, with each cell divided into $6$ linear tetrahedra. We refine the initial region where the bubble is located into two levels. After the refinement, the smallest element has a size of $0.025$ m, and the total number of elements is equal to $236,264$. The AMR/C procedure is applied every $4$ time steps and is based on the flux jump of the level-set function error, in which $h_{max}= 2$. Details about the AMR/C procedure may be found in \cite{rossa2013parallel, Grave2020}. We output a projected solution to a fixed mesh to enable the use of DMD as detailed in \cite{Barros2021}. The projected mesh comprises $768,000$ linear tetrahedral finite elements, with a finite element size of $0.025$ m. The time step size is defined as $\Delta t = 0.0125$ s, and we output the projected solutions using data compression at every $2$ time steps until the simulation reaches 3 seconds. PNG files are generated by Catalyst every time step. Therefore we have 120 snapshots and 240 images. The simulation runs in 8 cores.

\begin{figure}[ht!]
    \centering
    \includegraphics[width = 0.80\linewidth]{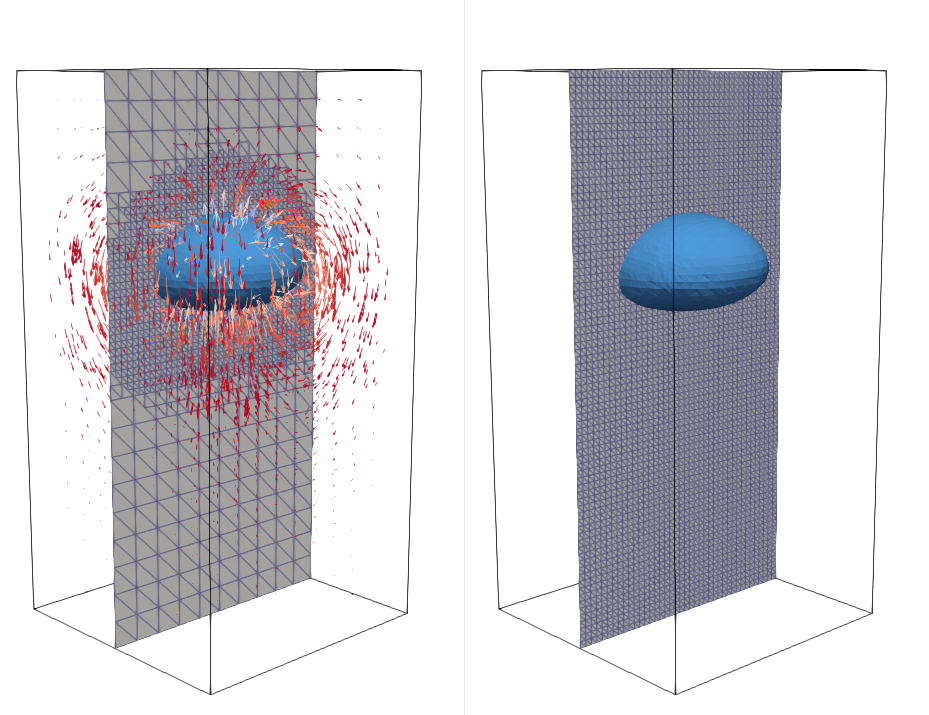}
    \caption{Bubble rising problem. View of bubble shape, velocity, and adapted mesh at $t=2.5$s. AMR/C solution (left) and projected mesh solution (right).
    }
    \label{fig:bubbleview}
\end{figure}

Again, the equations are spatially discretized using finite elements and RBVMS formulation. For time integration, the Backward-Euler is employed on the Navier-Stokes equations, and the BDF2 method is used for the convected level-set equation. The linear systems are solved utilizing GMRES(35) with Block-Jacobi preconditioner with ILU(0) within each block. We define the linear solver convergence's tolerance as equal to $10^{-6}$. The same tolerance is used to halt the nonlinear iterations. More details regarding the implementation can be seen on \cite{Grave2020, Barros2021}. Figure \ref{fig:bubbleview} left, shows the bubble shape, the velocity, and the adapted mesh at $t=2.5$s. We can see the solution projected in the reference mesh on the right. 


\par Data generated from the simulation follows the same workflow of the previous example. We test DMD on four cases: data with no compression, data with lossless compression, and data with lossy compression with two different accuracy thresholds. Table \ref{tab:data_bubble} shows the results in terms of storage requirements for the four cases. The lossless compression yields a $3.95\%$ reduction from the original data. Using lossy compression leads to more efficient results, being a $38.95\%$ reduction for the case where the threshold is $10^{-6}$ and $47.95\%$ for the case where the threshold is $10^{-3}$. We can also note in Table \ref{tab:data_bubble} that writing operations, generating the Catalyst images, and the mesh projection for enabling DMD on adapted meshes have a negligible contribution to the total execution time.

\begin{table}[!htp]\centering
\caption{Storage requirements for data generated on the rising bubble simulation.}
\begin{tabular}{l|cc|cccc}\toprule
                & \multicolumn{2}{c}{\textbf{Storage Required}} & \multicolumn{4}{c}{\textbf{Share of time spent on}}
                \\ & \textbf{XDMF/HDF5 (MB)} &  \textbf{Reduction (\%)}  &  \textbf{Solver} & \textbf{Write} & \textbf{Catalyst} & \textbf{Mesh Projection}   \\
                \midrule
No Compression          &855 &  --  &  89.80\%        & 0.61\%         &1.29\% &0.29\% \\
ZFP lossless            &824 & 3.63\%  & 89.45\%        & 0.65\%         &1.36\% &0.31\%  \\
ZFP accuracy $10^{-6}$  &522 & 38.95\% & 89.87\%        & 0.69\%         &1.30\% &0.29\%\\
ZFP accuracy $10^{-3}$  &445 & 47.95\% & 89.41\%        & 0.72\%      &1.37\% &0.31\%  \\
\bottomrule
\end{tabular}
\label{tab:data_bubble}
\end{table}

\par
\begin{figure}[ht!]
    \centering
    \includegraphics[width=0.49\linewidth]{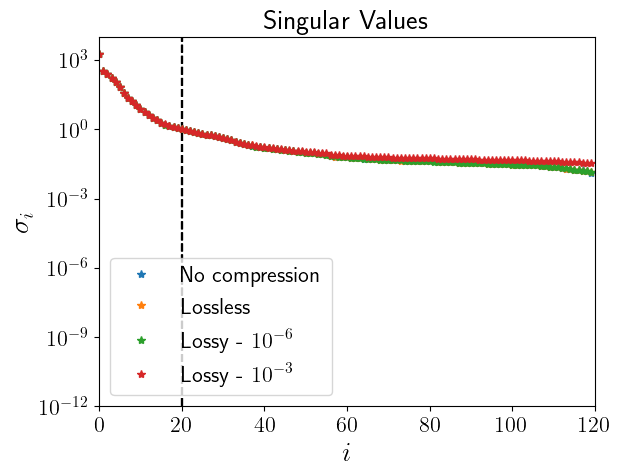}
    \includegraphics[width=0.49\linewidth]{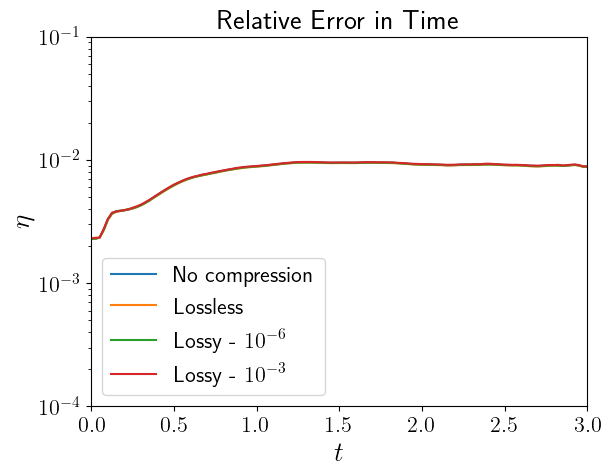}
    \caption{Singular values for the tested cases (left) and relative error in time for the reconstruction ($r=20$) (right).}
    \label{fig:dmd_bubble_comp}
\end{figure}

\begin{figure}[ht!]
    \centering
    \includegraphics[width=0.65\linewidth]{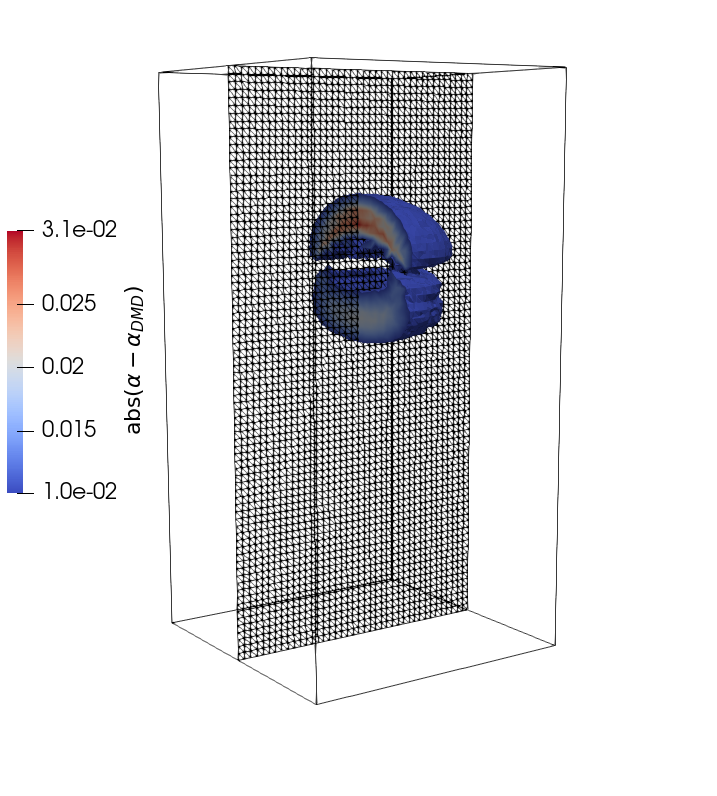}
    \caption{Results for the absolute error between DMD reconstruction with lossy compressed data (threshold $10^{-6}$) and original snapshot at $t = 2.5s$. The clipping shows the absolute error values larger than the threshold of $10^{-2}$. Note that the maximum absolute error is $3.1\times 10^{-2}$.}
    \label{fig:fom_dmd_bubble}
\end{figure}

 
We now proceed to apply DMD on the four cases shown in Table \ref{tab:data_bubble}. The DMD processing takes a few seconds to run. Figure \ref{fig:dmd_bubble_comp} shows the singular values decay of the snapshot matrix as well as the relative error in time between the numerical results and the respective approximation. We observe that the singular values' curves practically match, especially inside the truncation threshold for the DMD basis ($r = 20$, or 16,7\% of all snapshots available) represented by a vertical dashed line. The relative error for the four cases confirms that there is no significant difference when using data compression in this example. The curves illustrated in Figure \ref{fig:dmd_bubble_comp} show that the results are practically coincident except for the singular values tail for the lossy compression case with threshold $10^{-6}$ and do not depend on the data compression. Figure \ref{fig:fom_dmd_bubble} shows the results for the for the DMD approximation of the bubble interface (left) and the the absolute error (right) at $t = 2.5s$ obtained from ZFP compressed data with accuracy of $10^{-3}$. The results indicate good agreement between both. We also evaluate the Frobenius relative error between the snapshots matrix and the approximated solutions matrix in Table \ref{tab:frobenius_bubble}. We notice that the results are kept within an approximated Frobenius relative error of $10^{-3}$, indicating that the approximation results have good accuracy.

\begin{table}
    \centering
  \caption{Frobenius relative error between approximated and original snapshots for the bubble rising problem considering $r=20$}
    \begin{tabular}{cc}\toprule
        Test case & $\eta_F$ \\
        \midrule
        No compression &  $6.14 \times 10^{-3}$ \\
        Lossless       &  $6.14 \times 10^{-3}$ \\
        Lossy ($10^{-6}$) & $6.16 \times 10^{-3}$ \\
        Lossy ($10^{-3}$) & $7.15 \times 10^{-3}$ \\
        \bottomrule
    \end{tabular}
    \label{tab:frobenius_bubble}
\end{table}

\par
\begin{figure}[ht!]
    \centering
    \includegraphics[width=0.49\linewidth]{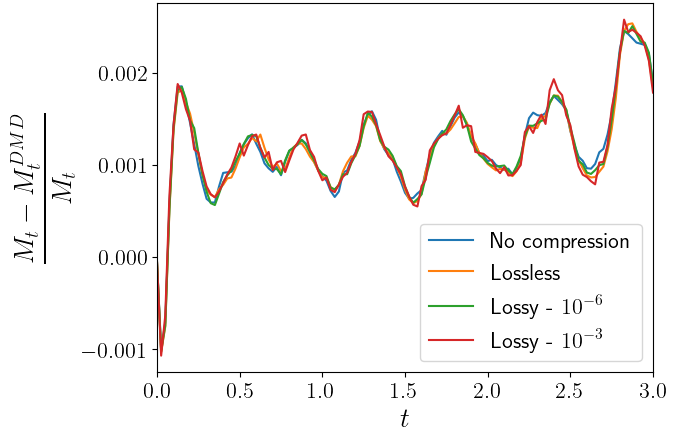}
    \caption{Time history of the bubble mass error for all cases.  For this problem, $M_t = \int_\Omega \alpha d\Omega$ and the superscript $^{DMD}$ refers to the same expression but taking the DMD reconstruction of $c$.}
    \label{fig:dmd_bubble_mass}
\end{figure}

 
We now evaluated if the data compression influences the reconstruction of the bubble mass. Figure \ref{fig:dmd_bubble_mass} shows the time history of the relative error of the bubble mass computed by the finite element simulation and the DMD reconstruction using uncompressed and compressed snapshots. We note that there is little difference between the time histories. Therefore, no physical information regarding the dynamics is lost using data compression when reconstructing derived quantities such as the bubble mass.

\subsection{DMD using streaming data}  
\par
\begin{figure}[ht!]
        \centering
        \includegraphics[trim={0cm 0cm 0cm 0cm},clip, width = 0.8\linewidth]{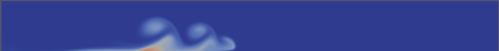}
        \includegraphics[trim={0cm 0cm 0cm 0cm},clip, width = 0.8\linewidth]{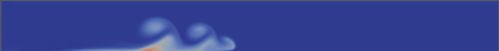}
        \caption{In-situ visualization concentration profile at the midplane image file at t=$12$s (top) and streaming DMD reconstructed image at the same time instant (bottom).}
        \label{fig:necker_snapshot}
    \end{figure}

 
In this section, we evaluate the use of streaming DMD on the images obtained from Paraview Catalyst generated in runtime as the simulations evolve. We consider images generated at the midplane of the 3D domains for both cases. 
For the lock-exchange simulation, we define $r=250$ as our threshold. For the bubble rising problem, given that we have 240 images, we take $r=60$, representing $25\%$ of the available data. In this study, we do not consider a DMD basis with less than $r$ vectors, although it would be possible to construct a basis with less than $r$ vectors for fewer snapshots. For instance, for $10$ relevant snapshots available, one could construct a temporary basis of $r = 10$ until reaching the threshold. In these examples, we start constructing the basis as soon as the first $r$ relevant snapshots are collected, so the DMD bases will always have their dimensions preserved while being updated with every new snapshot. In this case, however, the snapshots are pixel values of the PNG images instead of the nodal values obtained from the finite element discretization. Figure \ref{fig:necker_snapshot} shows the image produced with ParaView Catalyst for the concentration profile at the domain midplane in the lock-exchange problem and the streaming DMD reconstructed image at the same instant ($t = 12$s).


\begin{figure}[ht!]
    \centering
    \includegraphics[width=0.49\linewidth]{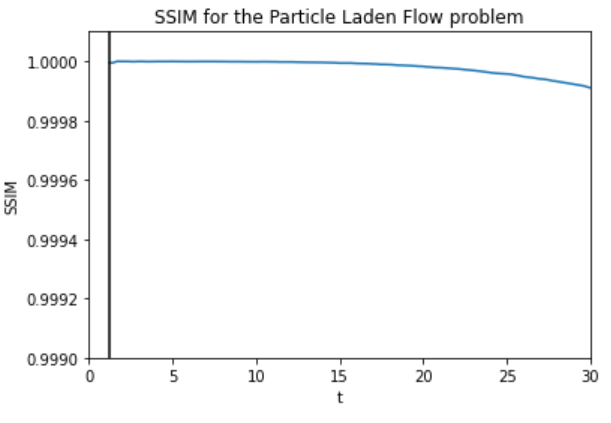}
    \caption{Evolution in time of the structural similarity index for the DMD reconstructed images for concentration at the midplane. }
    \label{fig:ssim_necker}
\end{figure}

For quantitative assessment of the reconstructed images, more dedicated metrics can be used to retrieve a straightforward and objective evaluation. In this study, we compute the Structural Similarity Index (SSIM)\cite{Wang2009, Wang2004, Avanaki2009} between output and input images to evaluate the approximation quality properly. The SSIM of a given group of pixels ranges from $-1$ (completely uncorrelated match) to $1$ (perfect match) and is a product of the computed contributions of luminance, contrast, and structure. The SSIM is computed as, 

\begin{equation}
    SSIM = \dfrac{(2\mu_{w_1}\mu_{w_2} + c_1)(2\sigma_{w_1w_2} + c_2)}{(\mu^2_{w_1} + \mu_{w_2}^2 + c_1)(\sigma_{w_1}^2 + \sigma_{w_2}^2 + c_2)}
\end{equation}

\noindent where the subscripts $w_1$ and $w_2$ are respective to two windows of common size of pixels, $\mu$ and $\sigma^2$ are, respectively, the average and variance of either $w_1$ and $w_2$, being $\sigma_{w_1w_2}$ the covariance of $w_1$ and $w_2$. In this study, we have used the SSIM implementation from scikit-image \cite{scikit-image}. The PNG files have $499 \times 51$ pixels, and the window size used is $7$ pixels - scikit-image`s default value. We can see in Figure \ref{fig:ssim_necker} that the metric stays close to 1.0 during the whole time interval of interest, indicating that the DMD reconstructed images for the concentration at the midplane are of good quality. 



\par
\begin{figure}[ht!]
        \centering
        \includegraphics[width = 0.49\linewidth]{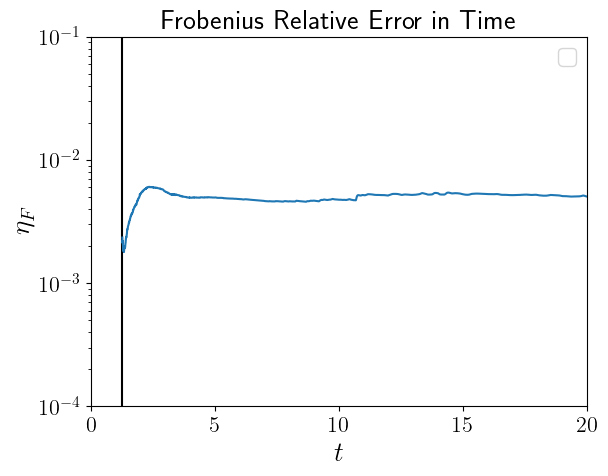}
        \includegraphics[width = 0.49\linewidth]{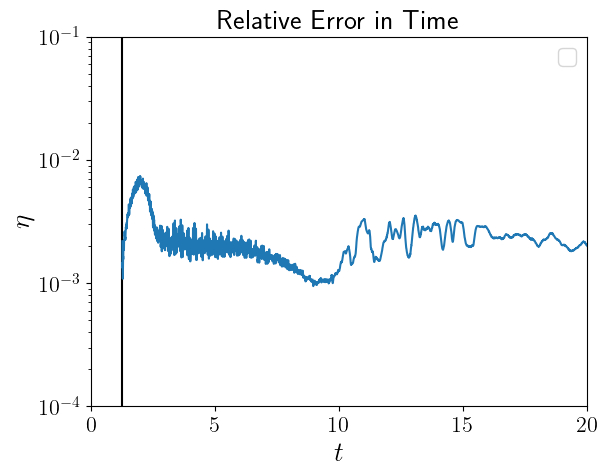}
        \caption{Evolution of the Frobenius relative error in time for ($r=250$) (left) and evolution of the $\mathcal{L}_2$ relative error in time for ($r=250$) (right) for the lock exchange problem.}
        \label{fig:evolution_frobenius_necker}
    \end{figure}

 We also evaluate the results in terms of Frobenius and $\mathcal{L}_2$ norms. Figure \ref{fig:evolution_frobenius_necker} shows the evolution in time for the Frobenius norm of the snapshots matrix and the matrix containing the approximations obtained from the streaming DMD algorithm. The $\mathcal{L}_2$ norm relative error can also be seen. The solid vertical line indicates the instant where the DMD basis has reached $r = 250$ vectors. From this point, the basis gets updated with the arrival of new snapshots relevant to the DMD modes but still preserves $r=250$. We observe that both metrics are below $10^{-2}$ for all time steps, indicating that the approximation is adequate. We also notice that the largest errors in Fig. \ref{fig:evolution_frobenius_necker} occur when the moving front is created (around $t = 3.5s$), meaning that the rapid physics during this stage are translated into more pronounced relative errors between the image files. We also notice this phenomenon in Fig. \ref{fig:ssim_necker}, being the smallest similarity index related to this stage.

\begin{figure}[ht!]
    \centering
    \includegraphics[width=0.49\linewidth]{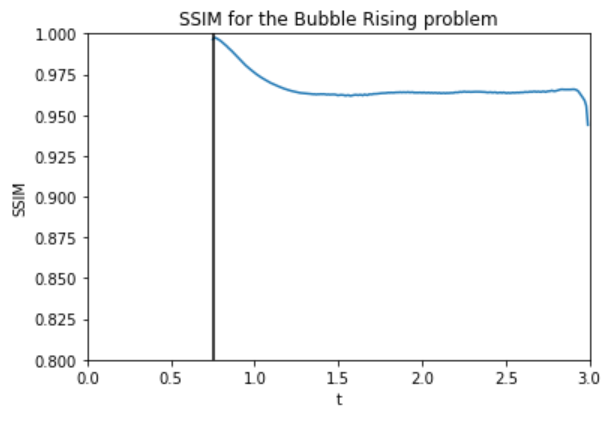}
    \caption{Evolution in time of the structural similarity index for the DMD reconstructed images for the bubble shape at the midplane. }
    \label{fig:ssim_bubble}
\end{figure}

\begin{figure}[ht!]
    \centering
    \includegraphics[width=.15\linewidth]{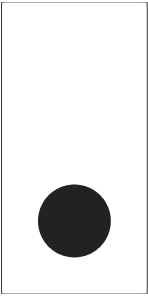}
    \includegraphics[width=.15\linewidth]{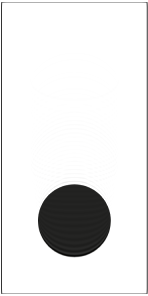}
    \hfill
    \includegraphics[width=.15\linewidth]{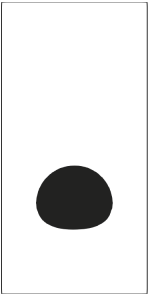}
    \includegraphics[width=.15\linewidth]{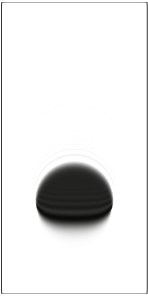}
    \hfill 
    \includegraphics[width=.15\linewidth]{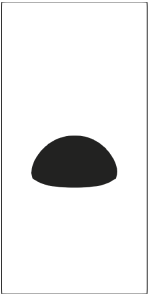}
    \includegraphics[width=.15\linewidth]{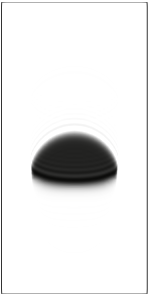}
    \\
    \includegraphics[width=.15\linewidth]{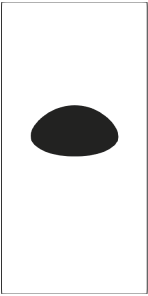}
    \includegraphics[width=.15\linewidth]{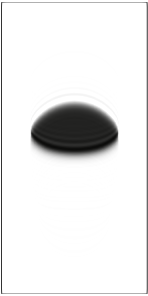}
    \hfill
    \includegraphics[width=.15\linewidth]{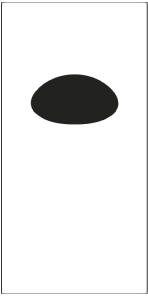}
    \includegraphics[width=.15\linewidth]{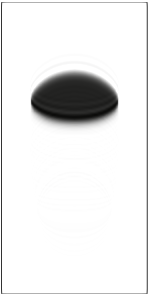}
    \hfill 
    \includegraphics[width=.15\linewidth]{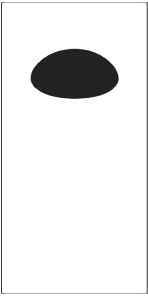}
    \includegraphics[width=.15\linewidth]{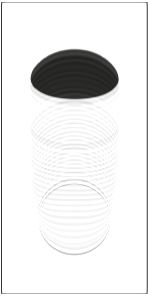}
    \caption{In-situ visualization image file of the bubble shape (left), streaming DMD reconstructed image ($r=60$) at the same instant (right). From top left to bottom right, $t = \{0.000, 0.625, 1.250, 1.875, 2.500, 3.000\}$s.}
    \label{fig:bubble_snapshot}
\end{figure}

For the bubble rising problem, the 2D images of the bubble shape at the midplane are presented in Figure \ref{fig:bubble_snapshot} for several time instants. The PNG files contain $148 \times 296$ pixels. The images produced by ParaView Catalyst can be seen as well as the DMD approximation at the same instants. Visually, the DMD reconstructions become blurry compared to the original image as the bubble moves towards the top of the domain. Also, for $t=2.4$s the DMD reconstruction presents some artifacts. This lack of sharpness is reflected in the evolution in time of the SSIM metric, as shown in Figure \ref{fig:ssim_bubble}. The SSIM metric decays immediately after the bubble starts to rise, stays at the same level until around $t=2.9$s, and is less accurate at later stages, where the bubble is near the top of the domain. However, note that most of the time, the SSIM is acceptable, above 0.95 \cite{nvidia-ssim}.

\begin{figure}[ht!]
    \centering
    \includegraphics[width = 0.49\linewidth]{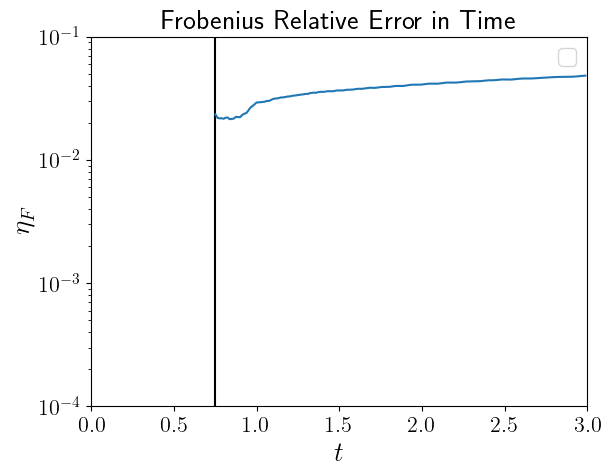}
    \includegraphics[width = 0.49\linewidth]{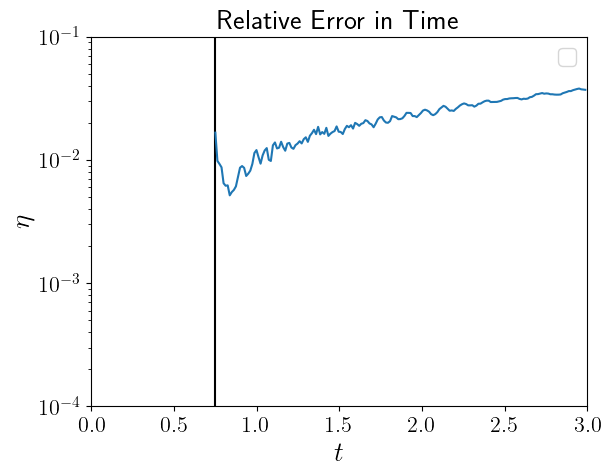}
    \caption{Evolution of the Frobenius relative error in time for ($r=60$) (left), evolution of the $\mathcal{L}_2$ relative error in time for ($r=60$) (right) for the bubble rising problem.}
    \label{fig:my_frobenius_bubble}
\end{figure}

Other metrics also show a moderate degradation of the quality of DMD reconstructions. Figure \ref{fig:my_frobenius_bubble} shows the time history of the Frobenius and $\mathcal{L}_2$ norms in time between approximation and snapshots. We can see in this Figure that as the bubble rises, both the relative Frobenius norm and the relative error grow, but both are still relatively low (less than $10^{-1}$).

\section{Conclusions}
\label{sec:conclusions}
This study proposes two approaches for enhancing the Dynamic Mode Decomposition workflow using techniques from the new Online Data Analysis and Reduction motif. First, we consider data compression in the snapshots. We use the standard DMD as our method of choice to reconstruct solutions to two highly nonlinear fluid dynamics problems: a particle-laden turbulent flow and a buoyancy-driven bubble rising problem. The nodal values of the output files of the simulations are stacked horizontally, creating a batch of snapshots (or a snapshots matrix) that will eventually be processed to generate the DMD basis. Large simulations may generate outputs with significant storage requirements, and data compression is one possibility to circumvent this issue. In this study, we use ZFP to reduce the size of the output files in the disk, yielding up to $50\%$ reduction on the original data with little loss in the DMD approximations. We observe that other quantities of interest, such as sediment and bubble mass, are well reproduced even when aggressive lossy compression is used. The reconstructed solutions reveal good accuracy with relative errors measured in the $\mathcal{L}_2$ norm less than $10^{-2}$. The second approach considers DMD reconstruction for images generated at simulation runtime. The Paraview Catalyst is used for this purpose. The PNG images are generated on-the-fly, and a different DMD algorithm is considered. The streaming DMD updates the DMD modes with the arrival of new snapshots instead of processing a batch of snapshots (or a snapshots matrix). The same fluid dynamics simulations are considered. In this case, the snapshots contain pixel values instead of nodal values. Results obtained for both problems using the Structural Similarity Index for comparison with the original Paraview Catalyst images show that the reconstructed images present good quality, particularly in the lock-exchange problem, where this metric is close to 1.0 during the time interval of interest. In the bubble rising problem, we observed that the Structural Similarity Index deteriorates as the bubble approaches the top of the domain. Nevertheless, the Structural Similarity Index is above 0.95. These low errors suggest that the DMD reconstructed images have the same quality as the original images. The relative error measured in the Frobenius and $\mathcal{L}_2$ norms, below $10^{-2}$ for both problems, also corroborates these findings.

Future works include extending data compression to other snapshots-based scientific machine learning methods for large-scale simulations. Another possible venue is using short-time streaming DMD predictions from in-situ visualization images to enhance online data analysis and simulation steering. As the simulation progresses, streaming DMD predictions can anticipate what will happen, opening the possibility to act to mitigate unforeseen problems while the simulation is still running. 

\section*{Acknowledgements}
This research was financed in part by the Coordena\c{c}\~ao de Aperfei\c{c}oamento de Pessoal de N\'ivel Superior - Brasil (CAPES) - Finance Code 001. This research has also received funding from CNPq, FAPERJ and FAPEMIG (APQ-01123-21). Computer time in Lobo Carneiro supercomputer was provided by the High Performance Computer Center at COPPE/Federal University of Rio de Janeiro, Brazil. 

%
%



\bibliographystyle{unsrt}  
\bibliography{references}  
\end{document}